\documentclass[pre,twocolumn,showpacs,preprintnumbers,floatfix,amsmath,amssymb,superscriptaddress]{revtex4}

\usepackage{graphicx}
\usepackage{latexsym}
\usepackage{amsmath}
\usepackage{amssymb}
\usepackage{amsfonts}
\usepackage{bm}

\advance\voffset 1.5cm

\newcommand{\la}{\left<}
\newcommand{\ra}{\right>}
\newcommand{\az}{\ensuremath{\approx 0}}
\newcommand{\qvec}{\ensuremath{\underline{q}}}
\newcommand{\rvec}{\ensuremath{\underline{r}}}

\newcommand{\uvec}{\ensuremath{\delta\underline{r}}}

\newcommand{\ddiff}{\ensuremath{\text{d}}}

\newcommand{\rijl}{\ensuremath{r_{l}}}
\newcommand{\sij}{\ensuremath{s_{l}}}
\newcommand{\sijl}{\ensuremath{s_{l}}}

\newcommand{\nx}{\ensuremath{n_{x}}}
\newcommand{\ny}{\ensuremath{n_{y}}}

\newcommand{\nijx}{\ensuremath{n_{x,l}}}
\newcommand{\nijy}{\ensuremath{n_{y,l}}}
\newcommand{\nija}{\ensuremath{n_{\alpha,l}}}
\newcommand{\nijb}{\ensuremath{n_{\beta,l}}}

\newcommand{\Tglass}{\mbox{$T_{\rm g}$}}
\newcommand{\kB}{\mbox{$k_{\rm B}$}}
\newcommand{\kBT}{\mbox{$k_{\rm B}T$}}

\newcommand{\Pid}{\ensuremath{P_\mathrm{id}}}
\newcommand{\Pidhat}{\ensuremath{\hat{P}_\mathrm{id}}}
\newcommand{\Pex}{\ensuremath{P_\mathrm{ex}}}
\newcommand{\Pexhat}{\ensuremath{\hat{P}_\mathrm{ex}}}

\newcommand{\sigexyy}{\ensuremath{\sigma^\mathrm{ex}_{yy}}}
\newcommand{\sigexaa}{\ensuremath{\sigma^\mathrm{ex}_{\alpha\alpha}}}
\newcommand{\sigexab}{\ensuremath{\sigma^\mathrm{ex}_{\alpha\beta}}}
\newcommand{\sighatexab}{\ensuremath{\hat{\sigma}^\mathrm{ex}_{\alpha\beta}}}
\newcommand{\sighatab}{\ensuremath{\hat{\sigma}_{\alpha\beta}}}
\newcommand{\sighatcd}{\ensuremath{\hat{\sigma}_{\gamma\delta}}}
\newcommand{\Fex}{\ensuremath{F_\mathrm{ex}}}
\newcommand{\Zex}{\ensuremath{Z_\mathrm{ex}}}
\newcommand{\Kid}{\ensuremath{K_\mathrm{id}}}
\newcommand{\Kex}{\ensuremath{K_\mathrm{ex}}}
\newcommand{\Uexs}{\ensuremath{U_s}}

\newcommand{\tauhat}{\ensuremath{\hat{\tau}}}
\newcommand{\gamhat}{\ensuremath{\hat{\gamma}}}
\newcommand{\dtauhat}{\ensuremath{\delta\hat{\tau}}}
\newcommand{\dgamhat}{\ensuremath{\delta\hat{\gamma}}}
\newcommand{\Xhat}{\ensuremath{\hat{X}}}
\newcommand{\Ihat}{\ensuremath{\hat{I}}}
\newcommand{\dXhat}{\ensuremath{\delta\hat{X}}}
\newcommand{\dIhat}{\ensuremath{\delta\hat{I}}}
\newcommand{\dAhat}{\ensuremath{\delta\hat{A}}}
\newcommand{\dBhat}{\ensuremath{\delta\hat{B}}}
\newcommand{\Vhat}{\ensuremath{\hat{V}}}
\newcommand{\Phat}{\ensuremath{\hat{P}}}
\newcommand{\dVhat}{\ensuremath{\delta\hat{V}}}
\newcommand{\dPhat}{\ensuremath{\delta\hat{P}}}

\newcommand{\uprime}{\ensuremath{u^{\prime}}}
\newcommand{\uprimeprime}{\ensuremath{u^{\prime\prime}}}
\newcommand{\utrunc}{\ensuremath{u_\mathrm{t}}}
\newcommand{\utruncprime}{\ensuremath{u^{\prime}_\mathrm{t}}}

\newcommand{\ushift}{\ensuremath{u_\mathrm{s}}}
\newcommand{\ushiftprime}{\ensuremath{u^{\prime}_\mathrm{s}}}
\newcommand{\ushiftprimeprime}{\ensuremath{u^{\prime\prime}_\mathrm{s}}}

\newcommand{\muA}{\ensuremath{\mu_\mathrm{A}}}
\newcommand{\muB}{\ensuremath{\mu_\mathrm{B}}}
\newcommand{\muBbare}{\ensuremath{\tilde{\mu}_\mathrm{B}}}
\newcommand{\muBcut}{\ensuremath{\Delta \mu_\mathrm{B}}}
\newcommand{\muF}{\ensuremath{\mu_\mathrm{F}}}
\newcommand{\muFtau}{\ensuremath{\left.\mu_\mathrm{F}}\right|_{\tau}}
\newcommand{\muFgam}{\ensuremath{\left.\mu_\mathrm{F}}\right|_{\gamma}}

\newcommand{\etaA}{\ensuremath{\eta_\mathrm{A}}}
\newcommand{\etaAid}{\ensuremath{\eta_\mathrm{A,id}}}
\newcommand{\etaAex}{\ensuremath{\eta_\mathrm{A,ex}}}
\newcommand{\etaB}{\ensuremath{\eta_\mathrm{B}}}
\newcommand{\etaF}{\ensuremath{\eta_\mathrm{F}}}
\newcommand{\etaFid}{\ensuremath{\eta_\mathrm{F,id}}}
\newcommand{\etaFex}{\ensuremath{\eta_\mathrm{F,ex}}}

\newcommand{\CFabcd}{\ensuremath{C_\mathrm{F}^{\alpha\beta\gamma\delta}}}
\newcommand{\Gmodgt}{\ensuremath{G_{\gamma\tau}}}
\newcommand{\Gmodgg}{\ensuremath{G_{\gamma\gamma}}}
\newcommand{\Gmodtt}{\ensuremath{G_{\tau\tau}}}
\newcommand{\Gmodttbare}{\ensuremath{\tilde{G}_{\tau\tau}}}
\newcommand{\cmodgt}{\ensuremath{c_{\gamma\tau}}}
\newcommand{\KmodVP}{\ensuremath{K_{vp}}}
\newcommand{\KmodVV}{\ensuremath{K_{vv}}}
\newcommand{\KmodPP}{\ensuremath{K_{pp}}}
\newcommand{\KmodPPbare}{\ensuremath{\tilde{K}_{pp}}}
\newcommand{\cmodVP}{\ensuremath{c_{vp}}}
\newcommand{\NVgT}{\ensuremath{\text{NV}\gamma\text{T}}}
\newcommand{\NPgT}{\ensuremath{\text{NP}\gamma\text{T}}}
\newcommand{\NVtT}{\ensuremath{\text{NV}\tau\text{T}}}
\newcommand{\NPtT}{\ensuremath{\text{NP}\tau\text{T}}}

\newcommand{\histoB}{\ensuremath{h_\mathrm{B}}}

\newcommand{\rcut}{\ensuremath{r_\mathrm{c}}}
\newcommand{\scut}{\ensuremath{s_\mathrm{c}}}
\newcommand{\scutm}{\ensuremath{s_\mathrm{c}^{-}}}
\newcommand{\smin}{\ensuremath{s_\mathrm{0}}}

\newcommand{\uLJ}{\ensuremath{u_\mathrm{LJ}}}

\newcommand{\nA}{\ensuremath{n_\mathrm{A}}}
\newcommand{\nB}{\ensuremath{n_\mathrm{B}}}

\newcommand{\epsAA}{\ensuremath{\epsilon_\mathrm{AA}}}
\newcommand{\epsBB}{\ensuremath{\epsilon_\mathrm{BB}}}
\newcommand{\epsAB}{\ensuremath{\epsilon_\mathrm{AB}}}
\newcommand{\epsij}{\ensuremath{\epsilon_{l}}}
\newcommand{\sigAA}{\ensuremath{\sigma_\mathrm{AA}}}
\newcommand{\sigBB}{\ensuremath{\sigma_\mathrm{BB}}}
\newcommand{\sigAB}{\ensuremath{\sigma_\mathrm{AB}}}

\newcommand{\sigmaij}{\ensuremath{\sigma_{l}}}
\newcommand{\drmax}{\ensuremath{\delta r_\text{max}}}
\newcommand{\Amax}{\ensuremath{a_\text{max}}}
\newcommand{\tauLJ}{\ensuremath{\tau_\mathrm{LJ}}}
\newcommand{\taurelax}{\ensuremath{\tau_\mathrm{r}}}
\newcommand{\Tstart}{\ensuremath{T_\mathrm{i}}}
\newcommand{\ttemp}{\ensuremath{t_\mathrm{eq}}}

\newcommand{\tmeas}{\ensuremath{t}}

\newcommand{\ca}{\ensuremath{c_{a}}}
\newcommand{\cb}{\ensuremath{c_{b}}}

\newcommand{\Ebarrier}{E_\text{b}}

\begin{document}

\title{Shear modulus of simulated glass-forming model systems:\\
Effects of boundary condition, temperature and sampling time}

\author{J.P.~Wittmer}
\email{joachim.wittmer@ics-cnrs.unistra.fr}
\affiliation{Institut Charles Sadron, Universit\'e de Strasbourg \& CNRS, 23 rue du Loess, 67034 Strasbourg Cedex, France}
\author{H.~Xu}
\affiliation{LCP-A2MC, Institut Jean Barriol, Universit\'e de Lorraine,\\ 1 bd Arago, 57078 Metz Cedex 03, France}
\author{P.~Poli\'nska}
\affiliation{Institut Charles Sadron, Universit\'e de Strasbourg \& CNRS, 23 rue du Loess, 67034 Strasbourg Cedex, France}
\author{F.~Weysser}
\affiliation{Institut Charles Sadron, Universit\'e de Strasbourg \& CNRS, 23 rue du Loess, 67034 Strasbourg Cedex, France}
\author{J. Baschnagel}
\affiliation{Institut Charles Sadron, Universit\'e de Strasbourg \& CNRS, 23 rue du Loess, 67034 Strasbourg Cedex, France}

\begin{abstract}
The shear modulus $G$ of two glass-forming colloidal model systems in $d=3$ and $d=2$ dimensions is investigated 
by means of, respectively, molecular dynamics and Monte Carlo simulations. Comparing ensembles where either the shear strain 
$\gamma$ or the conjugated (mean) shear stress $\tau$ are imposed, we compute $G$ from the respective stress 
and strain fluctuations as a function of temperature $T$ while keeping a constant normal pressure $P$.
The choice of the ensemble is seen to be highly relevant for the shear stress fluctuations $\muF(T)$
which at constant $\tau$ decay monotonously with $T$ following the affine shear elasticity $\muA(T)$,
i.e. a simple two-point correlation function.
At variance, non-monotonous behavior with a maximum at the glass transition temperature $\Tglass$
is demonstrated for $\muF(T)$ at constant $\gamma$. 
The increase of $G$  below $\Tglass$ is reasonably fitted for both models by a continuous cusp singularity,
$G(T) \propto (1-T/\Tglass)^{1/2}$, in qualitative agreement with some recent replica calculations. 
It is argued, however, that longer sampling times may lead to a sharper transition.
The additive jump discontinuity predicted by mode-coupling theory and other replica calculations
thus cannot ultimately be ruled out.
\end{abstract}

\date{\today}
\maketitle

\section{Introduction}
\label{sec_intro}
\paragraph*{Stress fluctuation formalism.}
Among the fundamental properties of any solid material are its elastic constants \cite{BornHuang,ChaikinBook,Wallace70}.
They describe the macroscopic response to weak external deformations and encode information about the
potential landscape of the system. 
In their seminal work \cite{Hoover69} Squire, Holt and Hoover derived an expression for the isothermal elastic constants 
extending the classical Born theory \cite{BornHuang} to canonical ensembles of imposed particle number $N$,
volume $V$ and {\em finite} (mean) temperature $T$. 
They found a correction term $\CFabcd$ to the Born expression 
for the elastic constants which involves the mean-square fluctuations of the stress \cite{Hoover69,Lutsko89,FrenkelSmitBook,SBM11}
\begin{equation}
\CFabcd \equiv - \beta V \left( \la \sighatab \sighatcd \ra - \la \sighatab \ra \la \sighatcd \ra \right )
\label{eq_CFabcd}
\end{equation}
with $\beta = 1/\kBT$ being the inverse temperature, $\kB$ Boltzmann's constant, 
$\sighatab$ the instantaneous value of a component of the stress tensor $\sigma_{\alpha\beta} = \langle \sighatab \rangle$
and Greek letters denoting the spatial coordinates in $d$ dimension. 
Perhaps due to the fact that the demonstration of the ``stress fluctuation formalism" in Ref.~\cite{Hoover69} 
assumed explicitly a well-defined reference position and a displacement field for the particles, 
it has not been appreciated that this approach is consistent (at least if the initial pressure is properly taken into account) 
with the well-known pressure fluctuation formula for the compression modulus $K$ of isotropic fluids 
\cite{RowlinsonBook,AllenTildesleyBook,FrenkelSmitBook,SBM11} 
which has been derived several decades earlier, presumably for the first time in the late 1940s by Rowlinson \cite{RowlinsonBook}. 
In any case, the stress fluctuation formalism provides a convenient route to calculating the elastic properties 
in computer simulations which has been widely used in the past
\cite{Ray85,Barrat88,FaKa00,WTBL02,TWLB02,TLWB04,LBTWB05,Lemaitre04,Barrat06,Pablo03,Pablo04,SBM11,SXM12}.
It has been generalized to systems with nonzero initial stress \cite{Ray85}, 
hard-sphere interactions \cite{FaKa00} and arbitrary continuous potentials \cite{Lutsko89}, 
or to the calculation of local mechanical properties \cite{Pablo04}.

\paragraph*{Low-temperature limit and non-affine displacements.}
In particular from Ref.~\cite{Lutsko89} it has become clear that the stress fluctuation correction 
to the leading Born term does not necessarily vanish in the zero-temperature limit. This is due to the fact
that when a system is subjected to a homogeneous deformation, the ensuing particle
displacements need not follow the macroscopic strain affinely \cite{WTBL02,TWLB02,TLWB04,LBTWB05,Barrat06}. 
The stress fluctuation term quantifies the extent of these non-affine displacements, 
whereas the Born term reflects the affine part of the particle displacements. 
How important the non-affine motions are, depends on the system under consideration \cite{Barrat06}. 
While the elastic properties of crystals with one atom per unit cell are given by the Born term only, 
stress fluctuations are significant for crystals with more complex unit cells \cite{Lutsko89}. 
They become particularly pronounced for polymer like soft materials \cite{SXM12} and amorphous solids 
\cite{Lemaitre04,Barrat06,Barrat88,WTBL02,TWLB02,TLWB04,LBTWB05,Pablo04,SBM11,LTWB06}.
This is revealed by recent simulation studies of various glass formers \cite{Barrat88}, 
like Lennard-Jones (LJ) mixtures \cite{WTBL02,TWLB02,TLWB04,LBTWB05},
polymer glasses \cite{Pablo04,SBM11} or silica melts \cite{LTWB06}.
Many of these simulation studies concentrate on the mechanical properties deep in the
glassy state, exploring for instance correlations between the non-affine displacement
field and vibrational anomalies of the glass (``Boson peak") \cite{LBTWB05,LTWB06},
the mechanical heterogeneity at the nanoscale \cite{Pablo04,Barrat09}
or the onset of molecular plasticity in the regime where the macroscopic deformation
is still elastic \cite{Barrat08}.

\begin{figure}[t]
\centerline{\resizebox{1.00\columnwidth}{!}{\includegraphics*{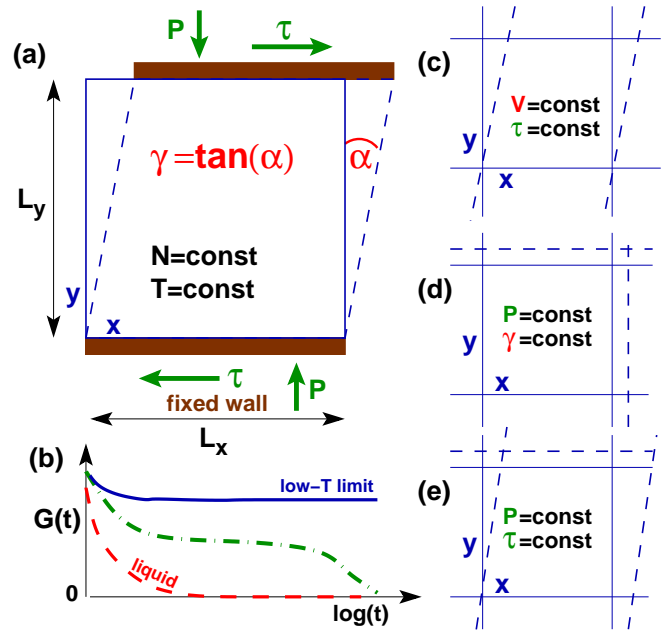}}}
\caption{Setup, notations and ensembles investigated numerically:
{\bf (a)} 
Schematic experimental setup for probing the shear modulus $G$. 
{\bf (b)}
Expected shear modulus $G(t)$ as a function of measurement time $t$ 
at $T \ll \Tglass$ (top curve), $T$ slightly below $\Tglass$
(middle curve) and in the liquid limit for $T \gg \Tglass$ (bottom curve). 
{\bf (c)}
We use periodic simulation boxes with deformable box shapes where we impose $\tau=0$. 
The shear modulus $G$ may be computed using Eq.~(\ref{eq_Gmodgt}) or Eq.~(\ref{eq_Gmodgg}) from the instantaneous 
strains $\gamhat$ and stresses $\tauhat$.
{\bf (d)}
Using Eq.~(\ref{eq_Gmodtt}) $G$ may also be computed 
using the affine shear elasticity $\muA$ and the shear stress fluctuations $\muF$ 
in ensembles of fixed strain ($\gamma =0$).
{\bf (e)}
Combined volumetric and shear strain fluctuations allowing the computation of both compression modulus $K$ 
and shear modulus $G$ from the respective strain fluctuations.
\label{fig_sketch}
}
\end{figure}

\paragraph*{Shear stress and shear modulus.}
Surprisingly, the temperature dependence of the elastic constants of glassy materials does not appear
to have been investigated extensively using the stress fluctuation formalism \cite{Barrat88,Pablo03,SBM11}. 
Following the pioneering work by Barrat {\em et al.} \cite{Barrat88}, we present in this work such a study 
where we focus on the behavior of the shear modulus $G$ of two isotropic colloidal glass-formers 
in $d=3$ and $d=2$ dimensions sampled, respectively, by means of molecular dynamics (MD) and 
Monte Carlo (MC) simulations \cite{AllenTildesleyBook,FrenkelSmitBook,LandauBinderBook}.
Since we are interested in isotropic systems one can avoid the inconvenient tensorial notation of the full
stress fluctuation formalism as shown in Fig.~\ref{fig_sketch}. Focusing on the shear in the $xy$-plane 
we use $\tau \equiv \sigma_{xy}$ for the mean shear stress and $\tauhat \equiv \hat{\sigma}_{xy}$ for
its instantaneous value. As an important contribution to the shear modulus we shall monitor the
shear stress fluctuations
\begin{equation}
\muF \equiv -C_\mathrm{F}^{xyxy} = \beta V \la \delta \tauhat^2 \ra  \ge 0
\label{eq_muF}
\end{equation}
as a function of temperature $T$. Quite generally, the shear modulus $G$ may 
then be computed according to the stress fluctuation formula \cite{Barrat88,SBM11} 
\begin{equation}
\Gmodtt \equiv \muA - \muF \label{eq_Gmodtt}
\end{equation}
where $\muA$ stands for the so-called ``affine shear elasticity" corresponding to an assumed {\em affine} 
response to an external shear strain $\gamma$. For pairwise interaction potentials it can be shown that
\begin{equation}
\muA  \equiv \muB - \Pex \ge 0, \label{eq_muAmuBPex}
\end{equation}
i.e. $\muA$ comprises the so-called ``Born-Lam\'e coefficient" $\muB$,
a simple moment of the first and the second derivative of the pair potential with respect
to the distance of the interacting particles \cite{BornHuang,Barrat06,SBM11},
and the excess contribution $\Pex$ to the total pressure $P$. 
These notations will be properly defined in Sec.~\ref{theo_stressfluctu} below where a simple derivation will be given. 
Since Eq.~(\ref{eq_Gmodtt}) is not the only operational definition of the shear modulus discussed in this work,
an index has been added reminding that the shear stress $\tauhat$ is used for the determination of $G$.

\paragraph*{Physical motivation and systems of interest.}
If an infinitesimal shear stress $\tau$ is applied at time $t=0$, an isotropic solid exhibits, quite generally, 
an elastic response quantified in strength by a finite static ($t$-independent) shear modulus ($G > 0$) 
as shown by the top curve in Fig.~\ref{fig_sketch}(b), whereas a liquid has a vanishing modulus ($G=0$) 
and flows on long time scales as shown by the bottom curve.
The shear modulus $G$ can thus serve as an ``order parameter" distinguishing 
liquid and solid states \cite{Zippelius06,Parisi10,Mezard10,Yoshino12,Szamel11,Klix12}. 
Upon melting, the shear modulus of crystalline solids is known to vanish discontinuously with $T$. 
A natural question which arises is that of the behavior of $G(T)$ for amorphous solids and glasses 
in the vicinity of the glass transition temperature $\Tglass$. 
Interestingly, qualitative different theoretical predictions have been put forward by mode-coupling theory (MCT)
\cite{GoetzeBook,Klix12} and some versions of replica theory \cite{Szamel11} predicting a {\em discontinuous jump} 
at the glass transition whereas other versions of replica theory \cite{Mezard10,Yoshino12} suggest a {\em continuous} 
transition in the static limit of large sampling times $t$. 
(See Ref.~\cite{Yoshino12} for a topical recent discussion of the replica approach.)
The numerical characterization of $G(T)$ for colloidal glass-former is thus of high current interest.
Since the stress fluctuation formalism can readily be added to the standard simulation codes of colloidal glass-formers 
\cite{Barrat88,Barrat06,LAMMPS}
at imposed particle number $N$, box volume $V$, box shape ($\gamma=0$) and temperature $T$, called \NVgT-ensembles,
this suggests the use of Eq.~(\ref{eq_Gmodtt}) to tackle this issue.
Note that the latter stress fluctuation relation should also be useful for wide range
of related systems, beyond the scope of the present work, including the glass transition of polymeric liquids \cite{SBM11}, 
colloidal gels \cite{Kob08} and self-assembled networks \cite{Zippelius06} 
as observed experimentally for hyperbranched polymer chains with sticky end-groups \cite{Friedrich10} or
bridged networks of telechelic polymers in water-oil emulsions \cite{Porte01b,Porte03}.

\paragraph*{Validity through a solid-liquid transition.}
This begs the delicate question of whether the general stress fluctuation formalism and especially Eq.~(\ref{eq_Gmodtt}) 
for the shear modulus only holds for elastic solids with a well-defined displacement field, 
as suggested in the early literature \cite{Hoover69,Lutsko89,FrenkelSmitBook}, 
or if the approach may actually be also applied {\em through} a solid-liquid transition
up to the liquid state \cite{RowlinsonBook,SXM12} with $\muF \to \muA$ and thus $G = \Gmodtt \to 0$ as it should for a liquid.
By reworking the theoretical arguments leading to Eq.~(\ref{eq_Gmodtt}) it will be seen that this
is in principal the case, at least if standard thermodynamics can be assumed to apply.
The latter point is indeed not self-evident for the colloidal glasses we are interested in where
{\em (i)} some degrees of freedom are frozen on the time scale probed and 
{\em (ii)} the measured response to an imposed infinitesimal shear $\tau$ may depend on the sampling time $t$, 
especially for temperatures $T$ around $\Tglass$ as shown in Fig.~\ref{fig_sketch}(b).
To be on the safe side, it thus appears to be appropriate to test numerically the applicability of the stress fluctuation 
formula, Eq.~(\ref{eq_Gmodtt}), using the corresponding strain fluctuation relations which are more fundamental 
on experimental grounds.

\paragraph*{Strain fluctuation relations.}
Panel (a) of Fig.~\ref{fig_sketch} shows a schematic setup for probing experimentally the shear modulus $G$
of isotropic systems confined between two rigid walls. We assume that the bottom wall is fixed and 
the particle number $N$, the (mean) normal pressure $P$ and the (mean) temperature $T$ are kept constant.
In linear response either a shear strain $\gamma = \tan(\alpha)$ or a (mean) shear stress $\tau$ may be imposed.
The thin solid line indicates the original unstrained reference system with $\gamma=0$ and $\tau=0$, 
the dashed line the deformed state. From both the mechanical and the thermodynamical point of view,
the shear modulus is defined as
\begin{equation}
G \equiv \left.\frac{\partial \tau(\gamma)}{\partial \gamma}\right|_{\gamma \to 0}
\label{eq_tGg}
\end{equation}
where the derivative may also be taken at $\tau=0$.
In practice, one obtains $G$ by fitting a linear slope to the $(\gamma,\tau)$-data at vanishing strain.
Obviously, the noise level is normally large for small strains. From the numerical point of view it would
thus be more appropriate to work at imposed mean stress $\tau=0$, to let the strain freely fluctuate and 
to sample the {\em instantaneous} values $(\gamhat,\tauhat)$. The shear modulus $G$ is then simply obtained 
by linear regression 
\begin{equation}
\Gmodgt \equiv \left.\frac{\la \delta \gamhat \delta \tauhat \ra}{\la  \delta \gamhat^2 \ra}\right|_{\tau=0}
\label{eq_Gmodgt}
\end{equation}
where the index indicates that both the instantaneous values $\gamhat$ and $\tauhat$ are used. 
(Note that imposing the strain $\gamma$ also fixes the instantaneous strain $\gamhat$ which thus cannot fluctuate.)
The associated dimensionless correlation coefficient
\begin{equation}
\cmodgt \equiv  \left.
\frac{\la \delta \gamhat \delta \tauhat \ra}{\sqrt{\la  \delta \gamhat^2 \ra \la  \delta \tauhat^2 \ra}}
\right|_{\tau=0}
\label{eq_cmodgt}
\end{equation}
allows to determine the quality of the fit. 
We do not know whether Eq.~(\ref{eq_Gmodgt}) and Eq.~(\ref{eq_cmodgt}) have actually been used in a real experiment,
however since no difficulty arises to work in a computer simulation at constant $\tau$ and to record $(\gamhat,\tauhat)$,
we use this linear regression as our key operational definition of $G$ which is used to judge the correctness
of the stress fluctuation relation, Eq.~(\ref{eq_Gmodtt}), computed at constant $\gamma$. 
To avoid additional physics at the walls, we use of course periodic boundary conditions \cite{AllenTildesleyBook} 
(the central image being sketched) as shown by the panels on the right-hand side of Fig.~\ref{fig_sketch}.
As discussed in Sec.~\ref{theo_Legendre}, it is possible to determine $\Gmodgt$ at constant volume (\NVtT-ensemble) 
as shown by panel (c) or constant normal pressure (\NPtT-ensemble) as shown by panel (e). The latter ensemble has 
the advantage that one can easily impose in addition the same normal pressure $P$ for all temperatures $T$ as it is justified 
for experimental reasons.
If thermodynamics holds, it is known (Sec.~\ref{theo_strainfluctu}) that
\begin{equation}
\left. \la \dgamhat \dtauhat \ra\right|_{\tau} = \kBT/V
\label{eq_dgamdtau}
\end{equation}
should hold \cite{foot_dgamdtau}. Using Eq.~(\ref{eq_Gmodgt}) this implies a second operational definition
\begin{equation}
\Gmodgg \equiv \left. \frac{\kBT / V}{\la \delta \gamhat^2 \ra} \right|_{\tau=0}
\label{eq_Gmodgg}
\end{equation}
which has the advantage that only the instantaneous shear strain $\gamhat$ has to be recorded.
The main technical point put forward in this paper is that all operational definitions yield similar results 
\begin{equation}
\left. \Gmodgt \right|_{\tau=0} \simeq \left. \Gmodgg \right|_{\tau=0} \approx \left. \Gmodtt \right|_{\gamma=0},
\label{eq_Gmods}
\end{equation}
even for temperatures where a strong dependence on the sampling time $t$ is observed.
In other words, Eq.~(\ref{eq_Gmods}) states that this time dependence applies similarly (albeit perhaps not exactly)
to the various moments computed for the three operational definitions of $G$ in different ensembles. 

\paragraph*{Outline.}
The paper is organized as follows:
Several theoretical issues are regrouped in Sec.~\ref{sec_theo}.
We begin the discussion in Sec.~\ref{theo_strainfluctu} by reminding a few useful general thermodynamic relations.
A simple derivation of the stress fluctuation formula, Eq.~(\ref{eq_Gmodtt}), for the
shear modulus $\Gmodtt$ in constant-$\gamma$ ensembles is given in Sec.~\ref{theo_stressfluctu}. 
(The analogous stress fluctuation formula for the compression modulus $K$ in constant-$V$ ensembles 
\cite{RowlinsonBook} is rederived in Appendix~\ref{app_Kstressfluctu}.)
Shear stress fluctuations $\muF$ in ensembles with different boundary conditions are discussed 
in Sec.~\ref{theo_Legendre}. 
For constant-$\tau$ ensembles $\muF = \muFtau$ will be shown to reduce to the affine shear elasticity $\muA$,
i.e. the functional $\Gmodtt$ must vanish for all temperatures. 
As pointed out recently \cite{XWP12}, the truncation of the interaction potentials used
in computational studies implies an impulsive correction for the affine shear elasticity $\muA$.
This is summarized in Sec.~\ref{theo_trunc}.
The numerical models used in this study are presented in Sec.~\ref{sec_algo}.
We turn then in Sec.~\ref{sec_res} to our main numerical results.
Starting with the high-$T$ liquid limit, Sec.~\ref{res_Thigh} shows that all our operational definitions 
correctly yield a vanishing shear modulus. 
Moving to the opposite low-$T$ limit we present in Sec.~\ref{res_dynmat} 
the elastic response for a topologically fixed network of harmonic springs constructed 
from the ``dynamical matrix" of our 2D model glass system \cite{WTBL02,TWLB02}. 
As shown in Sec.~\ref{res_Tlow}, the same behavior is found qualitatively for low-$T$ glassy bead systems.
The temperature dependence of stress fluctuations and shear moduli in different ensembles is then discussed in Sec.~\ref{res_Tvari}. 
As stated above, one central goal of this work is to determine the behavior of the shear modulus around $\Tglass$.
We conclude the paper in Sec.~\ref{sec_conc}. 
Thermodynamic relations and numerical findings related to the compression modulus $K$ are regrouped in Appendix~\ref{app_K}.
%

\section{Theoretical considerations}
\label{sec_theo}

\subsection{Reminder of thermodynamic relations}
\label{theo_strainfluctu}

We assume in the following that standard thermodynamics and thermostatistics \cite{Callen} 
can be applied and remind a few useful relations.
Let us denote an extensive variable by $X$ and its instantaneous value by $\Xhat$, 
the conjugated intensive variable by $I$ and its instantaneous value by $\Ihat$ \cite{foot_entropicintensive}.
With $F(T,X)$ being the free energy at temperature $T$ and imposed $X$, 
the intensive variable $I$ and the associated modulus $M$ are given by 
\begin{eqnarray}
I & \equiv & \frac{\partial F(T,X)}{\partial X}, \label{eq_FtoI} \\
M & \equiv & V \frac{\partial I }{\partial X} = V \frac{\partial^2 F(T,X)}{\partial X^2}. \label{eq_FtoM}
\end{eqnarray}
Alternatively, one may consider the Legendre transformation $H(T,I) \equiv F(T,X) - X I$ for which
\begin{eqnarray}
X & = & - \frac{\partial H(T,I)}{\partial I}, \label{eq_HtoI} \\
\frac{V}{M} & = & \frac{\partial X }{\partial I} = - \frac{\partial^2 H(T,I)}{\partial I^2}. \label{eq_HtoM}
\end{eqnarray} 
In our case, the extensive variable is $X= V \gamma$, the intensive variable 
$I = \tau$ \cite{foot_Ihat} and the modulus $M= G$ \cite{foot_Mbound}. 

We emphasize that there is an important difference between extensive and intensive variables:
If the extensive variable $X$ is fixed, $\Xhat$ cannot fluctuate. If, on the other hand,
the intensive variable $I = \langle \Ihat \rangle$ is imposed, not only $\Xhat$ but also $\Ihat$ 
may fluctuate \cite{foot_Tfluctu}.
In our case this means, that $\gamhat \equiv \gamma$ for the \NVgT-ensemble,
while the shear stress fluctuations $\muF$ defined above do clearly not vanish 
in the \NVtT-ensemble shown in Fig.~\ref{fig_sketch}(d).
Assuming $I$ to be imposed, it is well known that \cite{Callen}
\begin{eqnarray}
\left. \la \dXhat (\beta \dIhat) \ra\right|_{I} & = & 1 \label{eq_dXdI} \\
\left. \la \dXhat^2 \ra\right|_{I} & = & \frac{\partial X}{\partial {(\beta I)}} 
= \frac{V}{\beta M} \label{eq_dXdX}
\end{eqnarray}
with $\beta = 1/\kBT$ being the inverse temperature.
For our variables $X=V\gamma$ and $I=\tau$, Eq.~(\ref{eq_dXdI}) implies Eq.~(\ref{eq_dgamdtau}) 
from which it is seen that Eq.~(\ref{eq_Gmodgt}) is consistent with Eq.~(\ref{eq_Gmodgg}).
The latter relation is also obtained directly using Eq.~(\ref{eq_dXdX}).

As discussed in textbooks \cite{Callen,AllenTildesleyBook} a simple average $A = \langle \hat{A} \rangle$ 
of some observable ${\cal A}$ does not depend of the chosen ensemble, at least not if the system is large enough ($N,V \to \infty$).
A correlation function $\langle \dAhat \dBhat \rangle$ of two observables ${\cal A}$ and ${\cal B}$ may differ, however,
depending on whether $X$ or $I$ are imposed. As demonstrated by Lebowitz, Percus and Verlet in 1967 \cite{Lebowitz67}
one verifies that
\begin{equation}
\left. \la \dAhat \dBhat \ra\right|_{X} = \left. \la \dAhat \dBhat \ra\right|_{I}
- \frac{\partial (\beta I)}{\partial X} \frac{\partial A}{\partial (\beta I)} \frac{\partial B}{\partial (\beta I)}.
\label{eq_dAdB}
\end{equation}
If the extensive variable and at least one of the observables are identical, the left-hand side of Eq.~(\ref{eq_dAdB}) vanishes:
If $A=X$ and $B=I$, this yields Eq.~(\ref{eq_dXdI}). If $A=B=X$, this implies Eq.~(\ref{eq_dXdX}).
More importantly, for $A=B=I$ one obtains using Eq.~(\ref{eq_FtoM})
\begin{equation}
M = \beta V \left. \la \dIhat^2 \ra\right|_{I} - \beta V \left. \la \dIhat^2 \ra\right|_{X}.
\label{eq_dIdI}
\end{equation}
If it is possible to probe the intensive variable fluctuations in both ensembles,
this does in principle allow to determine the modulus $M$. 
(The latter equation might actually be taken as the definition of $M$ in cases where the thermodynamic reasoning becomes dubious.)
We emphasize that the modulus $M$ is an intrinsic property of the system and does {\em not} depend on the ensemble, 
albeit it is determined using ensemble depending correlation functions.
For a thermodynamically stable system, $M > 0$, Eq.~(\ref{eq_dIdI}) implies that 
\begin{equation}
\left.\la \dIhat^2 \ra\right|_{I} > \left.\la \dIhat^2 \ra\right|_{X}
\label{eq_IIgtIX}
\end{equation}
and that both fluctuations must become similar in the limit where the modulus $M$ becomes small. 
For the shear stress fluctuations $\muF$, Eq.~(\ref{eq_dIdI}) corresponds to the transformation
\begin{equation}
\left. \muF\right|_{V\gamma} = \muFtau - G,
\label{eq_muFtransform}
\end{equation}
and one thus expects $\muF$ to be larger for \NVtT-systems than for \NVgT-systems,
while in the liquid limit where $G \to 0$ the imposed boundary conditions should not matter. 
We shall verify Eq.~(\ref{eq_muFtransform}) numerically below.
A glance at the stress fluctuation formula, Eq.~(\ref{eq_Gmodtt}), suggests the identity 
\begin{equation}
\muFtau \stackrel{!}{=} \muA.
\label{eq_muAmuFtau}
\end{equation} 
Assuming Eq.~(\ref{eq_dgamdtau}) and Eq.~(\ref{eq_muAmuFtau}), it follows for the correlation coefficient 
$\cmodgt$ defined above that
\begin{equation}
\cmodgt = \sqrt{\Gmodgt/\muFtau} = \sqrt{G/\muA}.
\label{eq_cmodgt_muA}
\end{equation}
If the measured $(\gamhat,\tauhat)$ were indeed perfectly correlated, i.e. $\cmodgt =1$,
this implies $G = \muA$ and thus $\left.\muF\right|_{V\gamma} = 0$.
In fact, for all situations studied here we always have 
\begin{equation}
\cmodgt < 1 \mbox{, thus } G < \muA,
\label{eq_Gbound}
\end{equation}
i.e. the affine shear elasticity sets only an upper bound to the shear modulus
and a theory which only contains the affine strain response must {\em overpredict} $G$.
We show now directly that Eq.~(\ref{eq_Gmodtt}) and, hence, Eqs.~(\ref{eq_muAmuFtau},\ref{eq_cmodgt_muA},\ref{eq_Gbound}) 
indeed hold and this under quite general conditions.

\subsection{Shear modulus at imposed shear strain}
\label{theo_stressfluctu}

\paragraph*{Introduction.}
As shown in Fig.~\ref{fig_sketch}(a), we demonstrate Eq.~(\ref{eq_Gmodtt}) from the free energy change
associated with an imposed arbitrarily small pure shear strain $\gamma$ in the $xy$-plane 
assuming a  constant particle number $N$, a constant volume $V$ and a constant temperature $T$ (\NVgT-ensemble).
It is supposed that the total system Hamiltonian may be written as the sum of a kinetic energy
and a potential energy. Since for a plain shear strain at constant volume the ideal free energy
contribution does not change, i.e. is irrelevant for $\tau$ and $G$, we may focus on the excess 
free energy contribution 
\begin{equation}
\Fex(T,\gamma) = - \kBT \ln(\Zex(\gamma))
\label{eq_Fexgamma}
\end{equation}
due to the conservative interaction energy of the particles.
The (excess) partition function $\Zex(0)$ of the unperturbed system at $\gamma=0$ is the Boltzmann-weighted sum 
over all states $s$ of the system which are {\em accessible} within the measurement time $t$.
The argument $(0)$ is a short-hand for the unstrained reference.
Following the derivation of the compression modulus $K$ by Rowlinson \cite{RowlinsonBook} the partition function 
\begin{equation}
\Zex(\gamma) = \sum_s \exp(-\beta \Uexs(\gamma))
\label{eq_Zexgamma}
\end{equation}
of the sheared system is supposed to be the sum over the {\em same} states $s$, 
but with a different {\em metric} \cite{foot_metricchange} corresponding to the macroscopic strain
which changes the total interaction energy $\Uexs(\gamma)$ of 
state $s$ and, hence, the weight of the sheared configuration for the averages computed.
This is the central hypothesis made.
Interestingly, it is not necessary to specify explicitly the states of the unperturbed or perturbed system,
e.g., it is irrelevant whether the particles are distinguishable or not or whether they have a well-defined
reference position for defining a displacement field.

\paragraph*{Shear stress and modulus for general potential.}
Assuming Eq.~(\ref{eq_Fexgamma}) to hold we compute now the mean shear stress $\tau$ and the shear modulus $G$
for a general interaction potential $\Uexs(\gamma)$. We note for later convenience that
\begin{eqnarray}
\frac{\partial \ln(\Zex(\gamma))}{\partial \gamma} & = & \frac{\Zex^{\prime}(\gamma)}{\Zex(\gamma)} \label{eq_dlogZdgam} \\
\frac{\partial^2 \ln(\Zex(\gamma))}{\partial \gamma^2} & = & \frac{\Zex^{\prime\prime}(\gamma)}{\Zex(\gamma)}  
-\left(\frac{\Zex^{\prime}(\gamma)}{\Zex(\gamma)}\right)^2 \label{eq_ddlogZddgam}
\end{eqnarray}
for the derivatives of the free energy and
\begin{eqnarray}
\frac{\partial \Zex(\gamma)}{\partial \gamma} & = & - \sum_s \beta \Uexs^{\prime}(\gamma) \ e^{-\beta \Uexs(\gamma)} \label{eq_dZdgam} \\
\frac{\partial^2 \Zex(\gamma)}{\partial \gamma^2} & = 
& \sum_s \left( \beta \Uexs^{\prime}(\gamma) \right)^2 \ e^{-\beta \Uexs(\gamma)} \nonumber \\ 
& - & \sum_s \left( \beta \Uexs^{\prime\prime}(\gamma) \right) e^{-\beta \Uexs(\gamma)}
\label{eq_ddZddgam}
\end{eqnarray}
for the derivatives of the excess partition function 
where a prime denotes the derivative of a function $f(x)$ with respect to its argument $x$.
Using Eq.~(\ref{eq_dZdgam}) and taking finally the limit $\gamma \to 0$ one verifies for the shear stress that
\begin{equation}
\tau = \la \tauhat \ra \mbox{ with }  \tauhat \equiv 
\frac{1}{V} \left. \Uexs^{\prime}(\gamma)\right|_{\gamma=0}
\label{eq_tauhat_us}
\end{equation}
{\em defining} the instantaneous shear stress \cite{foot_Ihat}. Note that the average taken is defined as 
\begin{equation}
\langle \ldots \rangle = \frac{1}{\Zex(0)} \sum_s \ldots e^{-\beta \Uexs(0)}
\label{eq_average}
\end{equation}
using the weights of the unperturbed system. The shear stress thus measures the average change
of the total interaction energy $\Uexs(\gamma)$ taken at $\gamma=0$.
The shear modulus $G$ is obtained using in addition Eq.~(\ref{eq_ddlogZddgam}) and Eq.~(\ref{eq_ddZddgam})
and taking finally the  $\gamma \to 0$ limit. Confirming thus the stress fluctuation formula Eq.~(\ref{eq_Gmodtt}) 
stated in the Introduction, this yields
\begin{equation}
G = \Gmodtt \equiv \muA - \left.\muF\right|_{\gamma=0}
\label{eq_Gstressfluctu}
\end{equation}
with $\muA$ being the ``affine shear elasticity"
\begin{equation}
\muA \equiv \frac{1}{V} \la \left.\Uexs^{\prime\prime}(\gamma)\right|_{\gamma=0} \ra 
\label{eq_muAdef}
\end{equation}
already mentioned above and $\muF$ as defined in Eq.~(\ref{eq_muF}).
The comparison of Eq.~(\ref{eq_Gstressfluctu}) and Eq.~(\ref{eq_muFtransform})
confirms Eq.~(\ref{eq_muAmuFtau}).

\paragraph*{Comment.}
The affine shear elasticity $\muA$ corresponds to the change (second derivative) of the total energy 
which would be obtained if one actually strains {\em affinely} in a computer simulation
a given state $s$ without allowing the particles to relax their position.
As shown for athermal ($T\to 0$) amorphous bodies \cite{WTBL02,TWLB02,Barrat06},
the positions of the particles of such a strained configuration will of course 
in general change slightly to minimize the interaction energy relaxing thus the elastic moduli. 
This is also of relevance for thermalized solids where the non-affine displacements 
of the particles are driven by the minimization of the free energy.
It is for this reason that the shear-stress fluctuation term $\muF \ge 0$ must occur in Eq.~(\ref{eq_Gstressfluctu}) 
correcting the {\em overprediction} of the shear modulus $G$ by the affine shear elasticity $\muA$, Eq.~(\ref{eq_Gbound}).
This point has been overlooked in the early literature \cite{BornHuang}
and only appreciated much later \cite{RowlinsonBook,Hoover69,Lutsko88,Lutsko89,WTBL02,TWLB02} 
as discussed in Barrat's review \cite{Barrat06}.
Interestingly, as has been shown by Lutsko \cite{Lutsko89}, $\muF$ and other similarly defined
stress fluctuations become temperature independent in the harmonic ground state approximation for $T\to 0$.
Probing the stress fluctuations in a low-temperature simulation allows thus to determine the elastic moduli 
of athermal solids.

Please note that up to now we have not used explicitly the coordinate transformations
(metric change) associated with an affine shear strain. This is needed to obtain operational 
definitions for the instantaneous shear stress $\tauhat$ (and thus for $\tau$ and $\muF$) and 
the affine shear elasticity. 

\paragraph*{Coordinate transformation.}
As shown in Fig.~\ref{fig_sketch}(a), we assume a pure shear strain 
which transforms the $x$-coordinate of a particle position or 
the distance between two particles as
\begin{equation}
x(0) \Rightarrow x(\gamma) = x(0) + \gamma y(0)
\label{eq_strain_x}
\end{equation}
leaving all other coordinates unchanged \cite{foot_metricchange}. 
The squared distance $r^2$ between two particles thus transforms as
\begin{equation}
r^2(0) \Rightarrow r^2(\gamma) = r^2(0) + 2\gamma x(0) y(0) + \gamma^2 y^2(0)
\label{eq_strain_r}
\end{equation}
where we need to keep the $\gamma^2$-term for calculating the shear modulus below.
We note for later reference that this implies
\begin{eqnarray}
\frac{d r^2(\gamma)}{d\gamma} & = & 2 x(0) y(0) + 2 \gamma y^2(0) \label{eq_strain_rone} \\
\frac{d^2 r^2(\gamma)}{d\gamma^2} & = & 2 y^2(0) \label{eq_strain_rtwo} 
\end{eqnarray}
for, respectively, the first and the second derivative with respect to $\gamma$.
Let us now consider an arbitrary function $f(r(\gamma))$. 
Using Eq.~(\ref{eq_strain_rone}) it is readily seen that its first derivative with respect to $\gamma$
may be written as
\begin{eqnarray}
\frac{\partial f(r(\gamma))}{\partial \gamma} &  =  & 
f^{\prime}(r) \frac{1}{2r(0)} \left(2x(0) y(0) + 2 \gamma  y^2(0) \right) \label{eq_strain_froneA} \\
 & \Rightarrow & r f^{\prime}(r) \ \nx \ny \mbox{ for } \gamma \to 0. \label{eq_strain_froneB}
\end{eqnarray}
In the second step we have introduced the components $\nx = x(0)/r(0)$ and $\ny=y(0)/r(0)$ of the 
normalized distance vector between both particles. More generally, we denote by $n_{\alpha}$
the $\alpha$-component of a normalized vector with $\alpha = x, y, \ldots$
Stating only the small-$\gamma$ limit for the second derivative of $f(r(\gamma))$ with respect to $\gamma$ 
we note finally that
\begin{eqnarray}
\lim_{\gamma\to 0} \left( \frac{\partial^2 f(r(\gamma))}{\partial \gamma^2} \right) 
& = & \left(r^2 f^{\prime\prime}(r) - r f^{\prime}(r) \right) \ \nx^2 \ny^2 \nonumber \\
& + & r f^{\prime}(r) \  \ny \ny \label{eq_strain_frtwo}
\end{eqnarray} 
where we have dropped the argument $(0)$ for the distance $r$ of the unperturbed reference system.

\paragraph*{Pair interaction potentials.}
We assume now that the interaction energy $\Uexs(\gamma)$ of a configuration $s$  
is given by a pair interaction potential $u(r)$
between the particles \cite{FrenkelSmitBook,AllenTildesleyBook} with $r(\gamma)$ being the distance 
between two particles, i.e.
\begin{equation}
\Uexs(\gamma) = \sum_l u(\rijl(\gamma))
\label{eq_Upairinteraction}
\end{equation} 
where the index $l$ labels the interaction between the particles $i$ and $j$ with $i < j$.
Using the general result Eq.~(\ref{eq_tauhat_us}) for the shear stress it follows using 
Eq.~(\ref{eq_strain_froneA}) that 
\begin{equation}
\tauhat = \frac{1}{V} \sum_l \rijl u^{\prime}(\rijl) \ \nijx \ \nijy
\label{eq_tauhatKirkwood}
\end{equation}
where we have dropped the argument $(0)$ in the $\gamma \to 0$ limit.
We have thus rederived for the shear stress ($\alpha=x,\beta=y$) the well-known Kirkwood expression 
for the general excess stress tensor
\cite{AllenTildesleyBook}
\begin{equation}
\sigexab = \la \sighatexab \ra \mbox{ with } \sighatexab \equiv \frac{1}{V} \sum_l \rijl u^{\prime}(\rijl) \nija \nijb.
\label{eq_sigab}
\end{equation}
The affine shear elasticity $\muA$ is obtained using Eq.~(\ref{eq_muAdef}) and Eq.~(\ref{eq_strain_frtwo}).
This yields 
\begin{equation}
\muA = \muB + \sigexyy 
\label{eq_muAmuBsigyy}
\end{equation}
with the first contribution $\muB$ being the so-called ``Born-Lam\'e coefficient" \cite{BornHuang}
\begin{equation}
\muB \equiv \frac{1}{V} \la \sum_l \left(\rijl^2 u^{\prime\prime}(\rijl) - \rijl u^{\prime}(\rijl) \right) \nijx^2 \nijy^2 \ra.
\label{eq_muBdef}
\end{equation}
It corresponds to the first term in Eq.~(\ref{eq_strain_frtwo}).
The second contribution $\sigexyy$ stands for the normal excess stress in the $y$-direction,
i.e. the $\alpha=\beta=y$ component of the excess stress tensor. 

\paragraph*{Isotropic systems.}
In this work we focus on isotropic materials under isotropic external loads.
The normal (excess) stresses $\sigexaa$ for all directions $\alpha$ are thus identical, i.e.
\begin{equation}
\sigexaa \stackrel{!}{=} - \Pex = \frac{1}{dV} \la \sum_l \rijl u^{\prime}(\rijl) \ra
\label{eq_sigaaPex}
\end{equation}
with $\Pex \equiv P-\Pid$ being the excess pressure, $P$ the total pressure, 
$\Pid = \kBT \rho$ the ideal gas pressure and $d$ the spatial dimension. 
This implies that the affine shear elasticity is given by $\muA = \muB - \Pex$
as already stated in the Introduction, Eq.~(\ref{eq_muAmuBPex}). 
We note that for isotropic $d$-dimensional systems it can be shown \cite{SXM12,XWP12} 
that the Born-Lam\'e coefficient can be simplified as
\begin{equation}
\muB = 
\frac{1}{d(d+2) V} \sum_{l} \la \rijl^2 u^{\prime\prime}(\rijl) \underline{- \rijl u^{\prime}(\rijl)} \ra.
\label{eq_muB}
\end{equation}
The $d$-dependent prefactor stems from the assumed isotropy of the system
and the mathematical formula \cite{abramowitz}
\begin{equation}
\la \left( n_{\alpha} n_{\beta} \right)^2 \ra =
\frac{1}{d(d+2)} \left(1 + 2 \delta_{\alpha\beta} \right)
\label{eq_dprefactor}
\end{equation}
($\delta_{\alpha\beta}$ being the Kronecker symbol \cite{abramowitz})
for the components of a unit vector in $d$ dimensions pointing into arbitrary directions.
By comparing the excess pressure $\Pex$, Eq.~(\ref{eq_sigaaPex}), with the underlined second contribution
to the Born coefficient in Eq.~(\ref{eq_muB}) this implies that
\begin{equation}
\frac{-1}{d(d+2) V} \la \sum_{l} \rijl u^{\prime}(\rijl) \ra = \Pex/(d+2). 
\label{eq_Bornsecondterm}
\end{equation}
It is thus inconsistent to neglect the explicit excess pressure in Eq.~(\ref{eq_muAmuBPex})
while keeping the second term of the Born-Lam\'e coefficient $\muB$. This approximation
is only justified if the excess pressure $\Pex$ is negligible (and not the total pressure $P$).

\subsection{Shear stress fluctuations in different ensembles}
\label{theo_Legendre}

\paragraph*{Different ensembles.}
Being simple averages neither $\Pex$, $\muA$ or $\muB$ does depend for sufficiently large systems
on the chosen ensemble, i.e. irrespective of whether $\gamma$ or $\tau$ is imposed one expects
to obtain the same values. As already reminded in Sec.~\ref{theo_strainfluctu}, this is
different for fluctuations in general \cite{AllenTildesleyBook}. 
For instance, it is obviously pointless to use the shear-strain fluctuation formulae $\Gmodgt$ or $\Gmodgg$ 
in a constant-$\gamma$ ensemble. A more interesting result is predicted if $\Gmodtt \equiv \muA - \muF$ is 
computed in the ``wrong" constant-$\tau$ ensemble. Since $\left.\muF\right|_{\tau} = \muA$, 
Eq.~(\ref{eq_muFtransform}), one actually expects to observe $\Gmodtt = 0 $ for all temperatures $T$ if our thermodynamic 
reasoning holds through the glass transition up to the liquid state. 
We will test numerically this non-trivial prediction in Sec.~\ref{sec_res}.

Up to now we have only considered for simplicity of the presentation \NVgT- and \NVtT-ensembles at constant volume $V$.
Since on general grounds pure deviatoric and dilatational (volumetric) strains are decoupled (both strains commute), 
one expects the observable $\Gmodtt$ also to be applicable in the \NPgT-ensemble shown in Fig.~\ref{fig_sketch}(d)
and the observables $\Gmodgt$ and $\Gmodgg$ to be applicable in the \NPtT-ensemble as sketched in Fig.~\ref{fig_sketch}(e). 
In the latter ensemble one also expects to find $\Gmodtt = 0$ and $\left.\muF\right|_{\tau} = \muA$.
This will also be tested below.

\paragraph*{Direct derivation of $\muF$ for imposed $\tau$.}
We demonstrate now directly Eq.~(\ref{eq_muAmuFtau}) for the shear stress fluctuations
using the general mathematical identity
\begin{equation}
\la f(x) A(x) \ra = - \kBT \la A^{\prime}(x) \ra
\label{eq_intbyparts}
\end{equation}
with $x$ being an {\em unconstrained} coordinate, $A(x)$ some property, $f(x) \equiv - u^{\prime}(x)$ a ``force"
with respect to some ``energy" $u(x)$ and the average being Boltzmann weighted, 
i.e. $\la \ldots \ra \propto \int \ddiff x \ldots e^{-\beta u(x)}$. 
(It is assumed in Eq.~(\ref{eq_intbyparts}) that $u(x)$ decays sufficiently fast at the boundaries.)
Following work by Zwanzig \cite{Zwanzig65} we express first the instantaneous shear stress $\tauhat$
given above by the Kirkwood formula, Eq.~(\ref{eq_tauhatKirkwood}), by the alternative
virial representation \cite{AllenTildesleyBook}
\begin{equation}
\tauhat = \tau - \frac{1}{V} \sum_i y_i f_{x,i}.
\label{eq_tauhatfluctu}
\end{equation}
The second term, sometimes called the ``inner virial",
stands for a sum over all particles $i$ with $y_i$ being the $y$-coordinate of the particle
and $f_{x,i}$ the $x$-coordinate of the force acting on the particle. This contribution indeed vanishes
on average as can be seen using Eq.~(\ref{eq_intbyparts}),
\begin{equation}
\la y_i f_{x,i} \ra = - \kBT \la \frac{\partial y_i}{\partial x_i} \ra = 0,
\end{equation}
which shows that $\la \tauhat \ra = \tau$ as it must.
The stress fluctuations may thus be expressed as
\begin{eqnarray}
\muF & = & \beta V \la (\tauhat - \tau)^2 \ra \nonumber \\
     & = & \frac{\beta}{V} \la \sum_i y_i f_{x,i} \sum_j y_j f_{x,j} \ra \nonumber \\
     & = & - \frac{1}{V} \sum_{i} \la y_i \sum_j\frac{\partial y_j f_{x,j}}{\partial x_i} \ra 
\label{eq_stressfluctuvirial}
\end{eqnarray}
where we have used the integration by part, Eq.~(\ref{eq_tauhatfluctu}). 
This step requires that all particle positions are unconstrained and independent (generalized) coordinates.
Note that this is possible at imposed $\tau$, but cannot hold at fixed $\gamma$.
Assuming then pair interactions and using Eq.~(\ref{eq_tauhatKirkwood})
one can reformulate Eq.~(\ref{eq_stressfluctuvirial}) within a few lines. 
Without further approximation this yields 
\begin{equation}
\left. \muF\right|_{\tau} = - \frac{1}{V} \la \sum_l y_l^2 \frac{\partial f_{x,l}}{\partial x_l} \ra
\label{eq_stressfluctdecoupled}
\end{equation}
where the sum runs now over all interactions $l$ between particles $i < j$.
$x_l$ and $y_l$ refer to components of the distance vector of the interaction $l$
and $f_{x,l}$ to the $x$-component of the central force $f(\rijl) =-u^{\prime}(\rijl)$ 
between a particle pair at a distance $\rijl$. Note that in Eq.~(\ref{eq_stressfluctdecoupled})
the stress fluctuations stemming from different interactions are decoupled, i.e.
$\left.\muF\right|_{\tau}$ characterizes the self- or two-point contributions of directly interacting beads.
Since $f_x(r) = - u^{\prime}(r) x/r$ and, hence,
\begin{eqnarray}
- y^2 \frac{\partial f_x(r)}{\partial x} & = & \left( r^2 u^{\prime\prime}(r) - r^2 u^{\prime}(r)\right) \  \nx^2 \ny^2 \nonumber \\
& - & r u^{\prime}(r) \ \nx \ny, \label{eq_y2dfx}
\end{eqnarray}
this is identical to the affine shear elasticity $\muA$ derived at the end of Sec.~\ref{theo_stressfluctu}.
We have thus confirmed Eq.~(\ref{eq_muAmuFtau}).

Interestingly, the argument can be turned around:
\begin{enumerate}
\item[(1)] The stress fluctuation contribution term $\left. \muF\right|_{\tau} = \muA$ is obtained by 
the simple derivation given in this paragraph.
\item[(2)] Using the general Legendre transformation, Eq.~(\ref{eq_muFtransform}),
this implies $\Gmodtt = \muA -\muF$ for the \NVgT- and the \NPgT-ensemble.
\end{enumerate}
As already stressed, our thermodynamic derivation of Eq.~(\ref{eq_Gmodtt}) is rather general, not using in particular
a well-defined displacement field for the particle positions.

\subsection{Impulsive corrections for truncated potentials}
\label{theo_trunc}

\paragraph*{Truncation.}
It is common practice in computational condensed matter physics \cite{AllenTildesleyBook,FrenkelSmitBook}
to truncate a pair interaction potential $u(r)$, with $r$ being the distance between two particles $i$ and $j$,
at a conveniently chosen cutoff $\rcut$.
This allows to reduce the number of interactions to be computed and energy or force calculations become  
${\cal O}(N)$-processes. However, the truncation also introduces technical difficulties, e.g., 
instabilities in the numerical solution of differential equations
as investigated especially for the MD method \cite{AllenTildesleyBook,ToxvaerdDyre11}.
Without restricting much in practice the generality of our results, we assume below that
\begin{itemize}
\item
the pair potential $u(r)$ is short-ranged, i.e. that it decays within a few particle diameters,
\item
it  scales as $u(r) \equiv u(s)$ with the reduced dimensionless distance
$s=r/\sigmaij$ where $\sigmaij$ characterizes the length scale of the interaction $l$ and
\item
the same reduced cutoff $\scut = \rcut/\sigmaij$ is set for all interactions $l$.
\end{itemize}
For instance, for monodisperse particles with constant diameter $\sigma$,
as for the standard Lennard-Jones (LJ) potential \cite{AllenTildesleyBook},
\begin{equation}
\uLJ(s) = 4 \epsilon \left(\frac{1}{s^{12}} - \frac{1}{s^6} \right),
\label{eq_LJ}
\end{equation}
the scaling variable becomes $s=r/\sigma$ and the reduced cutoff $\scut = \rcut/\sigma$.
The effect of introducing $\scut$ is to replace $u(s)$ by the truncated potential
\begin{equation}
\utrunc(s) = u(s) H(\scut -s)
\label{eq_utrunc}
\end{equation}
with $H(s)$ being the Heaviside function \cite{abramowitz}.
Even if Eq.~(\ref{eq_utrunc}) is taken {\em by definition} as the new system Hamiltonian,
it is well known that impulsive corrections at the cutoff have to be taken into account in general
for the excess pressure $\Pex$ and other moments of the first derivatives of the potential
\cite{FrenkelSmitBook}. This is seen by considering the derivative of the truncated potential
\begin{equation}
\utruncprime(s) = \uprime(s) H(\scut -s) - u(s) \delta(s-\scut).
\label{eq_utruncprime}
\end{equation}

\paragraph*{Shifting of truncated potential.}
These corrections with respect to first derivatives can be avoided by considering a properly shifted potential 
\cite{FrenkelSmitBook}
\begin{equation}
\ushift(s) = \left( u(s) - u(\scut) \right) H(\scut -s)
\label{eq_ushift}
\end{equation}
since $\ushiftprime(s) = \uprime(s) H(\scut -s)$.
With this choice {\em no} impulsive corrections arise for moments of the instantaneous shear stress $\tauhat$ 
and of other components of the excess stress tensor $\sighatexab$, Eq.~(\ref{eq_sigab}).
Specifically, if the potential is shifted, all impulsive corrections are avoided for the
shear stress fluctuations $\muF$, Eq.~(\ref{eq_muF}).

\paragraph*{Correction to the Born coefficient.}
As shown in Ref.~\cite{XWP12}, the standard shifting of a truncated potential is, however, insufficient in general 
for properties involving second (and higher) derivatives of the potential. This is particulary the case for the 
Born coefficient $\muB$, Eq.~(\ref{eq_muB}), which is required to compute the affine shear elasticity $\muA$,
Eq.~(\ref{eq_muAmuBPex}).
The second derivative of the truncated and shifted potential being 
\begin{equation}
\ushiftprimeprime(s) =  \uprimeprime(s) H(\scut-s) - \uprime(s) \delta(s-\scut)
\label{eq_ushiftprimeprime}
\end{equation}
this implies that the Born coefficient reads $\muB = \muBbare - \muBcut$ with $\muBbare$ being the 
bare Born coefficient and $\muBcut$ the impulsive correction which needs to be taken into account.
The latter correction may be readily computed numerically from the configuration ensemble
using \cite{XWP12} 
\begin{eqnarray}
\muBcut & = & \lim_{s\to \scutm} \histoB(s) \mbox{ with } \nonumber\\
\histoB(s) & \equiv & \frac{1}{d (d+2) V} \sum_{l} \la  \sij^2 \uprime(\sij) \ \delta(\sij-s) \ra
\label{eq_histoLame}
\end{eqnarray}
being a weighted radial pair distribution function which is related to the standard radial pair 
distribution function $g(r)$ \cite{HansenBook}. 

\paragraph*{Mixtures and polydisperse systems.}
Below we shall consider model systems for mixtures and polydisperse systems where $\sigmaij$ may 
differ for each interaction $l$. For such mixed potentials $u(s)$, $\utrunc(s)$ and $\ushift(s)$ and 
their derivatives take in principal an explicit index $l$, i.e. one should write 
$u_l(s)$, $u_{\mathrm{t},l}(s)$, $u_{\mathrm{s},l}(s)$ and so on. 
(This is often not stated explicitly to keep a concise notation.)
For example one might wish to consider the generalization of the monodisperse LJ potential, Eq.~(\ref{eq_LJ}), 
to a mixture or polydisperse system with
\begin{equation}
u_l(s) = 4\epsij \left(s^{-12} - s^{-6} \right) \mbox{ with } s = r/\sigmaij
\label{eq_LJgeneralized}
\end{equation}
where $\epsij$ and $\sigmaij$ are fixed for each interaction $l$.
In practice, each particle $i$ may be characterized by an energy scale $E_i$ and a ``diameter" $D_i$.
The interaction parameters $\epsij(E_i,E_j)$ and $\sigmaij(D_i,D_j)$ are then given in terms of
specified functions of these properties \cite{TWLB02}. 
The extensively studied Kob-Andersen (KA) model for binary mixtures of beads of type A and B \cite{Kob95},
is a particular case of Eq.~(\ref{eq_LJgeneralized}) with fixed interaction ranges $\sigAA$, $\sigBB$ and $\sigAB$ and
energy parameters $\epsAA$, $\epsBB$ and $\epsAB$ characterizing, respectively,
AA-, BB- and AB-contacts.
The expression Eq.~(\ref{eq_histoLame}) remains valid for such explicitly $l$-dependent potentials.

\paragraph*{Shear modulus and compression modulus.}
Since $\Gmodtt = \muB - \Pex -\muF$, Eq.~(\ref{eq_histoLame}) implies in turn 
\begin{equation}
\Gmodtt = \Gmodttbare -\muBcut
\label{eq_Gmodttcorrected}
\end{equation}
with $\Gmodttbare$ being the uncorrected bare shear modulus.
As we shall see in Sec.~\ref{res_Thigh}, the correction $\muBcut$ is of importance for the {\em precise} determination
of $\Gmodtt$ close to the glass transition where the shear modulus must vanish.
An impulsive correction has also to be taken into account for the compression modulus $K=\KmodPP$
computed from the normal pressure fluctuations in a constant volume ensemble as discussed in Appendix~\ref{app_K}.
Using the symmetry of isotropic systems, Eq.~(\ref{eq_etaBmuB}), one verifies \cite{XWP12}
\begin{equation}
\KmodPP = \KmodPPbare- \frac{2+d}{d} \muBcut
\label{eq_KmodPPcorrected}
\end{equation}
with $\KmodPPbare$ being the uncorrected compression modulus.

\section{Algorithmic and technical issues}
\label{sec_algo}

\begin{figure}[t]
\centerline{\resizebox{1.0\columnwidth}{!}{\includegraphics*{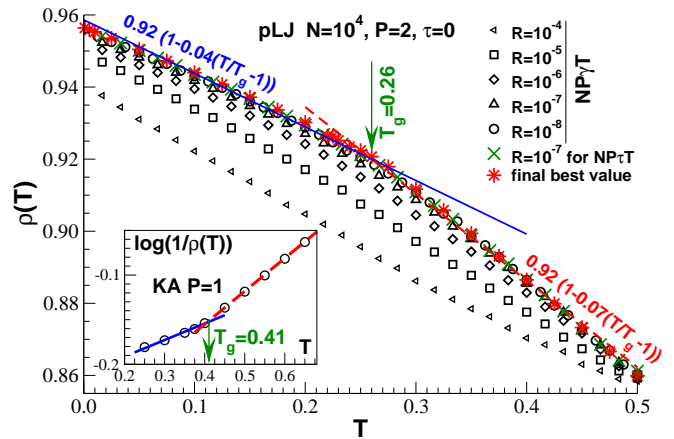}}}
\caption{Density $\rho(T)$ and determination of glass transition temperature $\Tglass$.
Inset:
Plotting $\log(1/\rho)$ {\em vs.} $T$ for the KA model as suggested in Ref.~\cite{SBM11}
reveals a linear low-$T$ (solid line) and a linear high-$T$ regime which confirms $\Tglass \approx 0.41$ \cite{Kob95}.
Main panel:
Number density $\rho$ for pLJ systems at constant normal pressure $P=2$
as a function of temperature $T$ for different quench rates $R$ in units of MCS.
The crosses refer to a quench in the \NPtT-ensemble where the box shape is allowed to fluctuate
at $\tau=0$. All other data refer to the \NPgT-ensemble with $\gamma=0$.
The solid line and the dashed line are linear fits to the best density estimates (stars)
obtained for, respectively, low and high temperatures. By matching both lines
one determines $\Tglass \approx 0.26$.
\label{fig_quench_rho}
}
\end{figure}

\paragraph*{Introduction.}
The computational data presented in this work have been obtained using two extremely well studied models of colloidal glass-formers 
which are described in detail elsewhere \cite{Kob95,WTBL02,TWLB02,Barrat06}. We present first both models and the molecular dynamics (MD) 
and Monte Carlo (MC) \cite{AllenTildesleyBook,FrenkelSmitBook,LandauBinderBook} methods used to compute them in the different ensembles
(\NVgT, \NVtT, \NPgT \ and \NPtT) compared in this study.
We comment then on the quench protocol and locate the glass transition temperature $\Tglass$ for both models via dilatometry, 
as shown in Fig.~\ref{fig_quench_rho}. This is needed as a prerequisite to our study of the elastic behavior in Sec.~\ref{sec_res}.
As sketched in Fig.~\ref{fig_sketch}, periodic boundary conditions are applied in all spatial directions 
with $L_x=L_y=\ldots$ for the linear dimensions of the $d$-dimensional simulation box.
Time scales are measured in units of the LJ time $\tauLJ = (m\sigma^2/\epsilon)^{1/2}$ \cite{AllenTildesleyBook} 
for our MD simulations ($m$ being the particle mass, $\sigma$ the reference length scale and $\epsilon$ the
LJ energy scale) and in units of Monte Carlo Steps (MCS) for our MC simulations.
LJ units are used throughout this work ($\epsilon=\sigma=m=1$) and Boltzmann's constant $\kB$ is also set to unity.

\paragraph*{Kob-Andersen model.}
The Kob-Andersen (KA) model \cite{Kob95} for binary mixtures of LJ beads in $d=3$ dimensions has been investigated 
by means of MD simulation  \cite{FrenkelSmitBook} taking advantage of the public domain LAMMPS implementation \cite{LAMMPS}.
We use $N=\nA+\nB=6912$ beads per simulation box and molar fractions $\ca=\nA/n=0.8$ and $\cb=\nB/n=0.2$ 
for both types of beads $A$ and $B$. Following Ref.~\cite{Kob95} we set $\sigAA=1.0\sigma$, $\sigBB=0.88 \sigma$ and $\sigAB=0.8 \sigma$
for the interaction range and $\epsAA=1.0 \epsilon$, $\epsBB=0.5\epsilon$ and $\epsAB=1.5 \epsilon$
for the LJ energy scales. Only data for the usual (reduced) cutoff $\scut=\rcut/\sigmaij =2.5$ are presented.
For the \NPgT-ensemble we use the Nos\'e-Hoover barostat provided by the LAMMPS code (``fix npt command") \cite{LAMMPS}
which is used to impose a constant pressure $P=1$ for all temperatures. Starting with an equilibrated configuration
at $\Tglass=0.7$ well above the glass transition temperature $\Tglass = 0.41$ (as confirmed in the inset of Fig.~\ref{fig_quench_rho}), 
the system is slowly quenched in the \NPgT-ensemble.
The imposed mean temperature $T(t)$ varies linearly as
$T(t) = \Tstart - R t$ with $R=5\times 10^{-5}$ 
being the constant quench rate. After a tempering time $\ttemp = 10^4$ we compute over a sampling time $t= 5\cdot 10^4$ various 
properties for the isobaric system at fixed temperature such as the moduli $\Gmodtt$ represented by the filled spheres in Fig.~\ref{fig_GT}. 
Fixing only then the volume (\NVgT-ensemble) the system is again tempered ($\ttemp = 10^4$) and various properties 
as the ones recorded in Table~\ref{tab_KA} are computed using a sampling time $t= 5\cdot 10^4$.
Note that in the \NVgT-ensemble the equations of motion are integrated using a velocity Verlet algorithm \cite{AllenTildesleyBook} 
with a time step $\delta t = 0.01$ and systems are kept at the imposed temperature $T$ using a Langevin thermostat \cite{AllenTildesleyBook} 
with friction coefficient $\gamma=0.5$. 
By redoing for several temperatures around and below $\Tglass$ the tempering and the production runs we have checked that the presented 
results remain unchanged (within numerical accuracy) and that ageing effects are absolutely negligible. 
Averages are taken over four independent configurations. As one may expect from the discussion in Sec.~\ref{theo_Legendre}, 
very similar results have been found for both ensembles for simple averages, for instance for the affine shear elasticity $\muA$, 
and for the shear modulus $\Gmodtt$, Eq.~(\ref{eq_Gmodtt}).
Unfortunately, we are yet unable to present data for the KA model data at imposed shear stress $\tau$ (\NVtT-, \NPtT-ensembles).

\begin{table}[t]
\begin{tabular}{|c||c|c|c|c|c|c|c|c|c|c|c|}
\hline
$T$                   &
$\rho$                & 
$e \rho$              & 
$\muBcut$             &
$\muB$                &
$\muA$                &
$\Gmodtt$              &
$\etaB$               &
$\etaA$               &
$\KmodPP$              
\\ \hline
0.025 & 1.24 & -9.42 & 0.80 & 52.1 & 51.15 &21.9 & 86.2 & 87.2 & 82.4  \\
0.100 & 1.20 & -8.85 & 0.80 & 50.4 & 49.51 &19.6 & 83.4 & 84.5 & 75.7  \\
0.200 & 1.21 & -8.57 & 0.78 & 48.3 & 47.52 &17.3 & 80.0 & 81.2 & 67.7  \\
0.250 & 1.20 & -8.31 & 0.77 & 47.2 & 46.54 &13.9 & 78.2 & 79.6 & 63.1  \\
0.310 & 1.19 & -7.99 & 0.74 & 45.7 & 45.10 &11.8 & 75.8 & 77.2 & 56.9  \\
0.350 & 1.18 & -7.82 & 0.73 & 45.1 & 44.45 &9.3  & 73.9 & 76.1 & 52.6 \\
0.375 & 1.17 & -7.67 & 0.72 & 44.6 & 43.96 &6.1  & 73.9 & 75.3 & 50.5 \\
0.400 & 1.17 & -7.49 & 0.71 & 43.8 & 43.24 &2.5  & 72.6 & 74.1 & 45.3 \\
0.450 & 1.15 & -7.07 & 0.68 & 42.0 & 41.48 &0.03 & 69.6 & 71.1 & 37.9 \\
0.800 & 1.00 & -4.49 & 0.55 & 29.7 & 29.5  &\az  & 49.4 & 51.2 & 16.6  \\
\hline
\end{tabular}
\vspace*{0.5cm}
\caption[]{Some properties for the KA model at normal pressure $P=1$ and shear stress $\tau=0$:
the mean density $\rho$, 
the mean energy per volume $e\rho$,
the impulsive correction term $\muBcut$ (Sec.~\ref{theo_trunc}),
the (corrected) Born-Lam\'e coefficient $\muB$,
the affine shear elasticity $\muA= \muB -\Pex$,
the shear modulus $\Gmodtt$ obtained in the \NVgT-ensemble,
the hypervirial $\etaB$ and the affine dilatation elasticity $\etaA= 2 \Pid + \etaB + \Pex$ discussed 
in Appendix~\ref{app_K} and the compression modulus $\KmodPP$
obtained for the \NVgT-ensemble.
The affine elasticities $\muA$ and $\etaA$ are the natural scales for,
respectively, the shear modulus $G$ and the compression modulus $K$.
\label{tab_KA}}
\end{table}

\paragraph*{Polydisperse LJ beads.}
A systematic comparison of constant-$\gamma$ and constant-$\tau$ ensembles has been performed, however, by MC simulation of a specific 
case of the generalized LJ potential, Eq.~(\ref{eq_LJgeneralized}), where all interaction energies are identical, $\epsij= \epsilon$, 
and the interaction range is set by the Lorentz rule $\sigmaij=(D_i+D_j)/2$ \cite{HansenBook} with $D_i$ and $D_j$ being the diameters 
of the interacting particles. Only the strictly two-dimensional (2D) case ($d=2$) is considered.
Following Ref.~\cite{TWLB02}, the bead diameters of this polydisperse LJ (pLJ) model
are uniformly distributed between $0.8\sigma$ and $1.2\sigma$.  
For the examples reported in Sec.~\ref{sec_res} we have used $N=10000$ beads per box.
We only present data for a reduced cutoff $\scut=2.0\smin$ with $\smin = 2^{1/6}$ being the
the minimum of the polydisperse LJ potential, Eq.~(\ref{eq_LJgeneralized}).
Various properties obtained for the pLJ model are summarized in Table~\ref{tab_pLJ}.

\begin{table}[t]
\begin{tabular}{|c||c|c|c|c|c|c|c|c|c|c|c|}
\hline
$T$                   &
$\rho$                & 
$e \rho$              & 
$\muBcut$             &
$\muA$                &
$\Gmodgt$             &
$\Gmodgg$             &
$\Gmodtt$             &
$\etaA$               &
$\KmodVP$             &
$\KmodPP$              
\\ \hline
0.001 & 0.96 & -2.70 & 0.31 & 33.9 &15.6 & 15.7 &15.5& 71.7 & 70.8 & 70.5  \\
0.010 & 0.96 & -2.69 & 0.38 & 33.7 &14.3 & 14.2 &13.0& 71.3 & 69.7 & 69.7 \\
0.100 & 0.94 & -2.60 & 0.39 & 32.4 &10.9 & 10.9 &10.3& 68.8 & 61.2 & 61.4 \\
0.200 & 0.93 & -2.48 & 0.38 & 31.1 &7.2  & 7.2  &7.4 & 66.2 & 51.2 & 50.6\\
0.225 & 0.93 & -2.45 & 0.38 & 30.6 &4.6  & 4.6  &3.7 & 65.3 & 47.9 & 46.4\\
0.250 & 0.92 & -2.42 & 0.37 & 30.0 &1.4  & 1.4  &0.94& 64.0 & 41.0 & 44.4\\
0.275 & 0.92 & -2.38 & 0.37 & 29.7 &0.6  & 0.6  &0.05& 63.4 & 40.0 & 39.4\\
0.300 & 0.91 & -2.33 & 0.35 & 29.3 &0.1  & 0.1  &0.09& 62.6 & 37.3 & 37.6\\
0.325 & 0.91 & -2.29 & 0.39 & 28.9 &\az  &\az   &\az & 61.7 & 34.0 & 34.3\\
0.350 & 0.90 & -2.24 & 0.33 & 28.3 &\az  &\az   &\az & 60.5 & 31.6 & 31.2\\
1.000 & 0.73 & -1.23 & 0.22 & 17.0 &\az  &\az   &\az & 38.0 & 9.3  & 9.0  \\
\hline
\end{tabular}
\vspace*{0.5cm}
\caption[]{Some properties of the pLJ model at $P=2$ and $\tau=0$:
the mean density $\rho$, 
the mean energy per volume $e \rho$,
the impulsive correction term $\muBcut$,
the affine shear elasticity $\muA$,
the shear moduli $\Gmodgt$ and $\Gmodgg$ obtained in the \NPtT-ensemble and
$\Gmodtt$ obtained in the \NVgT-ensemble,
the affine dilatation elasticity $\etaA$,
the compression modulus $\KmodVP$ obtained in the \NPtT-ensemble
and the Rowlinson formula $\KmodPP$ for the \NVgT-ensemble.
All data have been obtained for a sampling time $t = 10^{7}$ MCS.
Please note that the values given correspond to the best values obtained with the most appropriate technique and 
may differ thus slightly from the data represented in the figures for a specific protocol.
\label{tab_pLJ}}
\end{table}

\begin{figure}[t]
\centerline{\resizebox{1.0\columnwidth}{!}{\includegraphics*{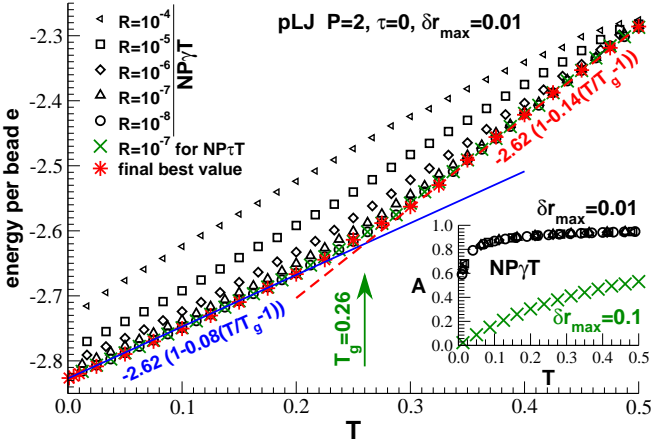}}}
\caption{Energy per particle $e$ (main panel) and acceptance rate $A$ (inset) obtained for pLJ systems quenched at constant 
pressure $P=2$ at a given quench rate $R$ as indicated.  The crosses refer to a quench at imposed $\tau=0$.
The solid line and the dashed line are linear fits to the energy for, respectively, low and high temperatures
confirming the value $\Tglass \approx 0.26$ for the pLJ model.
Only local MC moves with one fixed (non-adaptive) maximum particle displacement $\drmax$ for all temperatures $T$ have been 
used for the MC data shown in the main panel. The acceptance rate decreases thus with decreasing $T$ as shown in the inset  
for $\drmax=0.01$ (upper data) and $\drmax=0.1$.
\label{fig_quench_e_pLJ}
}
\end{figure}

\paragraph*{Local MC moves for the pLJ model.}
For all ensembles considered here local MC moves are used (albeit not exclusively) where a particle is chosen randomly
and a displacement $\uvec$, uniformly distributed over a disk of radius $\drmax$, from the current position $\rvec$
of the particle is attempted. The corresponding energy change $\delta E$ is calculated and the move is accepted
using the standard Metropolis criterion \cite{LandauBinderBook}.
The maxium displacement distance $\drmax$ is chosen such that the acceptance rate $A$ remains reasonable, 
i.e. $0.1 \le A \le 0.5$ \cite{LandauBinderBook}.
As may be seen from the inset in Fig.~\ref{fig_quench_e_pLJ} where the acceptance rate $A(T)$ is shown as a
function of temperature $T$, we find that temperatures below $T=0.1$ are best sampled using $\drmax \approx 0.01$,
whereas $\drmax \approx 0.1$ is a good choice for the interesting temperature regime between $T=0.2$ and $T=0.3$ around 
the glass transition temperature $\Tglass = 0.26$. 
(See the main panel of Fig.~\ref{fig_quench_rho} for the determination of $\Tglass$.)
Various data sets are presented in Sec.~\ref{sec_res} for a fixed value $\drmax=0.1$ which allows a reasonable comparison of the 
sampling time dependence for temperatures around the glass transition. This value is not necessarily the best choice for tempering 
the system and for computing static equilibrium properties at the given temperature, especially below $T \approx 0.2$.

\paragraph*{Collective plane-wave MC moves.}
Local MC jumps in the \NVgT-ensemble become obviously inefficient for relaxing large scale structural properties \cite{LandauBinderBook}
for the lowest temperatures computed for the pLJ model ($T \le 0.1$) as may be seen, e.g., from the finite, albeit very small shear stresses $\tau$ 
which happen to occur naturally in some of the quenched configurations (especially if the quench rate $R$ is too large)
and can only with difficulty be relaxed using local jumps {\em alone} \cite{foot_localonly}.
We have thus crosschecked and improved the values obtained using only local MC moves by adding global MC move attempts
for all particles with longitudinal and (more importantly) transverse plane waves commensurate with the simulation box. 
The random amplitudes and phases of the waves are assumed to be uniformly distributed.
Note that the maximum amplitude $\Amax(q)$ of each wave is chosen inversely proportional to the length $q=|\qvec|$ of the wavevector 
in agreement with continuum theory. This allows to keep the same acceptance rate for each global mode if $q \ll 1$.
As one expects, the plane-wave displacement attempts becomes inappropriate for large $q$ where continuum elasticity 
breaks down \cite{WTBL02,TWLB02,Barrat06} and, hence, the acceptance rate $A(q)$ gets too large to be efficient \cite{foot_adaptive}.
Note that the maximum amplitude $\Amax(q)$ has to be chosen much smaller for the longitudinal waves as for the transverse ones,
i.e. since the systems are essentially incompressible, the transverse modes naturally are more effective for quenching 
and equilibrating the systems. 
These global plane-wave moves have been added for the tempering of all presented configurations below $T=0.3$ with $\ttemp=10^7$
and for production runs over sampling times $t=10^7$ for static properties for $T < 0.2$. 
Details concerning the collective plane-wave MC moves must be given elsewhere.

\paragraph*{MC sampling of different ensembles.}
In this paper we shall be concerned more with the consequences of affine displacements associated with collective MC moves 
changing the overall box volume $V$ and shear strain $\gamma$. Using again a Metropolis criterion for accepting a suggested
change of $V$ or $\gamma$, these pure dilatational and pure shear strain fluctuations are used to impose a mean normal pressure 
$P=2$ and a mean shear stress $\tau=0$. 
For a shear strain altering MC move one first chooses randomly a strain fluctuation $\delta \gamma$ with $|\delta \gamma| \le \delta \gamma_\mathrm{max}$.
Assuming an imposed shear stress $\tau$ the suggested $\delta \gamma$ is accepted if
\begin{equation}
\xi \le \exp\left(-\beta \delta E + \beta \delta \gamma \tau \right)
\label{eq_timposed}
\end{equation} 
with $\xi$ being a uniformly distributed random variable with $0 \le \xi < 1$.
The energy change $\delta E$ associated with the affine strain may be calculated using Eq.~(\ref{eq_strain_r}).
Since $\tau =0$ is supposed in this study, the last term in the exponential drops out.
(As noted in Sec.~\ref{theo_stressfluctu}, the volume remains unchanged in a pure shear strain
and, hence, no translational entropy contribution appears.) 
The maximum shear strain step $\delta \gamma_\mathrm{max}$ is fixed such that at the given temperature 
the acceptance rate remains reasonable \cite{foot_maxvalues}. 
As described in more detail in Ref.~\cite{AllenTildesleyBook}, one similarly chooses a relative volume change $\delta V/V$ 
with $V$ being the current volume \cite{foot_maxvalues}. The volume altering move is accepted at an imposed pressure $P$ if
\begin{equation}
\xi \le \exp\left(-\beta \delta E + N \log(1 + \delta V/V) - \beta \delta V P  \right)
\label{eq_Pimposed}
\end{equation} 
where $\delta E$ may be computed using Eq.~(\ref{eq_strain_r_epsilon}).
The logarithmic contribution corresponds to the change of the translational entropy.
Using Eq.~(\ref{eq_timposed}) after each MCS for local MC moves (and global plane-wave MC moves if these are added)
realizes the \NVtT-ensemble shown in Fig.~\ref{fig_sketch}(c), 
using Eq.~(\ref{eq_Pimposed}) the \NPgT-ensemble shown in Fig.~\ref{fig_sketch}(d) and 
using in turn both strain fluctuations the \NPtT-ensemble in Fig.~\ref{fig_sketch}(e).
We remind that \NVgT- and \NPgT-ensembles are used to determine $\Gmodtt$
and \NVtT- and \NPtT-ensembles to obtain $\Gmodgt$ and $\Gmodgg$.
As shown in Fig.~\ref{fig_quench_rho}, we quench the configurations starting from $\Tstart=0.5$ 
in the \NPgT- and the \NPtT-ensemble using a constant quench rate $R$. Note that the smallest
quench rate $R=10^{-8}$ for one configuration in the \NPgT-shown did require alone a run over six months
using one core of an Intel Xeon E5410 processor. Interestingly, similar results are obtained with the \NPtT-ensemble
only using a rate $R=10^{-7}$. (Due to the additional computation for the shear strain fluctuations this is, however,
only a factor five faster.)
The configurations created by the \NPgT-quenches are used (after tempering) for the sampling in the \NVgT- and \NPgT-ensembles,
the configurations obtained by the \NPtT-quenches for the sampling in the \NVtT- and \NPtT-ensembles.
Only two independent configurations have been sampled following the full protocol for each temperature and each of the four ensembles.
Instead of increasing further the number of configurations (which will be done in the future) we have focused in the present
preliminary study on long tempering times and production runs over $t=10^7$ MCS which have been redone several times
for a few temperatures (summing up to total runs of $\approx 5\cdot 10^7$ MCS for $T=0.2$ and $T=0.25$) 
to check for equilibration problems and to verify that ageing effects can be ignored. 
We firmly believe that the breaking of the time translational invariance is not a relevant issue for
the strong sampling time dependence reported below for the shear moduli.

\paragraph*{Glass transition temperature.}
\label{algo_sample}
The inset of Fig.~\ref{fig_quench_rho} presents for the KA model how the glass transition temperature $\Tglass$
may be obtained using standard dilatometry where the rescaled density $\log(1/\rho)$ is traced as a function of temperature $T$
as in Ref.~\cite{SBM11} for a similar polymer glass problem. This reveals a linear low-$T$ (solid line) and a linear high-$T$ 
regime. The intercept of both lines confirms the well-known value $\Tglass \approx 0.41$ from the literature \cite{Kob95}.
The main panel presents the unscaled number density $\rho$ for pLJ systems at constant normal pressure $P=2$
as a function of temperature $T$ for different quench rates $R$ in MCS.
Only local MC moves with a fixed maximum particle displacement $\drmax=0.01$ for all temperatures $T$ are used for the
examples given. The crosses refer to a quench in the \NPtT-ensemble where the box shape is allowed to fluctuate at $\tau=0$. 
All other data refer to the \NPgT-ensemble with $\gamma=0$. The solid line and the dashed line are linear fits to the best 
density estimates indicated by the stars obtained for, respectively, low and high temperatures. By matching both lines
one determines $\Tglass \approx 0.26$. A similar value is given if $\log(1/\rho)$ is plotted as a function of $T$.
The main panel of Fig.~\ref{fig_quench_e_pLJ} shows the interaction energy per particle $e$ of the pLJ model
as a function of temperature for the same quench protocols as in Fig.~\ref{fig_quench_rho}. 
The solid line and the dashed line are linear fits to the energy for, respectively, low and high temperatures.
This confirms $\Tglass \approx 0.26$ for the pLJ model \cite{foot_dynamics}.  
%

\section{Computational results}
\label{sec_res}

\begin{figure}[t]
\centerline{\resizebox{1.0\columnwidth}{!}{\includegraphics*{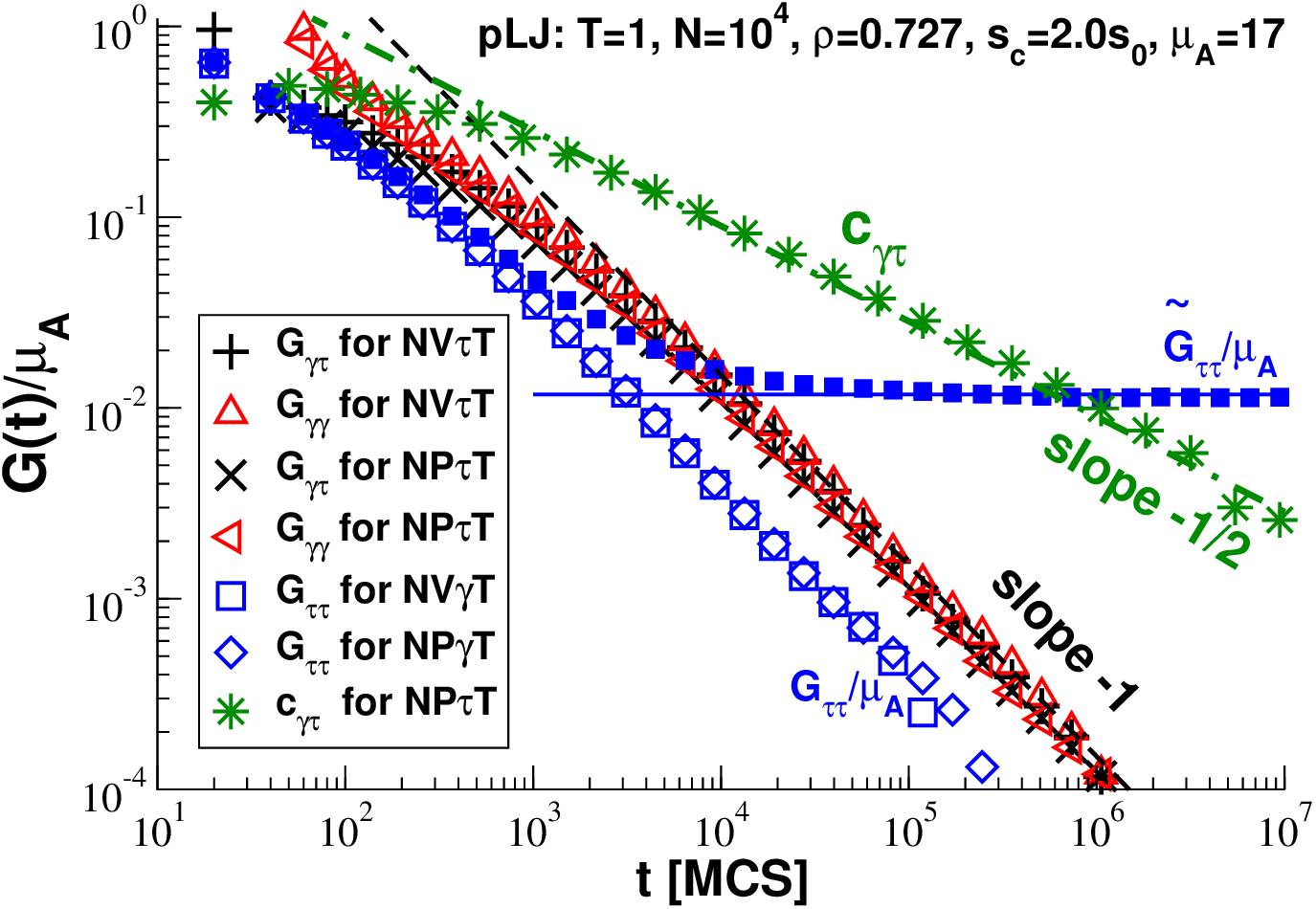}}}
\caption{Rescaled shear modulus $G(t)/\muA$ {\em vs.} sampling time $t$ in the high-$T$ liquid limit
of the pLJ model comparing various ensembles and observables. 
The filled squares show the bare shear modulus $\Gmodttbare$ obtained using Eq.~(\ref{eq_Gmodtt}) 
for the \NVgT-ensemble {\em without} the impulsive correction $\muBcut \approx 0.2$ (horizontal line). 
The corrected modulus $\Gmodtt=\Gmodttbare-\muBcut$ for the \NVgT-ensemble (open squares) and 
the \NPgT-ensemble (open diamonds) vanishes as it should.
Note that $\Gmodtt$ decreases slightly faster than $\Gmodgt\approx \Gmodgg$. 
As shown by the dashed line, all operational definitions of $G$ decay inversely with $t$.
The dash-dotted line indicates the expected power-law exponent $-1/2$ for the correlation coefficient $\cmodgt$ (stars).
\label{fig_Gt_Thigh}
}
\end{figure}

\subsection{High-temperature liquid limit}
\label{res_Thigh}

We begin our discussion by focusing on the high temperature liquid limit where all 
reasonable operational definitions of the shear modulus $G$ must vanish.
The results presented in Fig.~\ref{fig_Gt_Thigh} have been obtained for the pLJ model at mean pressure $P=2$,
mean shear stress $\tau = 0$ and mean temperature $T=1$, i.e. far above the glass transition temperature $\Tglass \approx 0.26$. 
Only local MC monomer displacements with a maximum jump displacement $\drmax =0.1$ are reported here.
Instantaneous properties relevant for the moments are written down every $10$ MCS and averaged using standard 
gliding averages \cite{AllenTildesleyBook}, i.e. we compute mean values and fluctuations for a given time interval
$[t_0,t_1=t_0+t]$ and average over all possible intervals of length $t$.
The horizontal axis indicates the latter interval length $t$ in units of MCS.
The vertical axis is made dimensionless by rescaling the moduli with the (corrected) affine shear elasticity 
$\muA = \muB - \Pex \approx 17$ from Table~\ref{tab_pLJ}. (Being a simple average, $\muA$ is found to be 
identical for all computed ensembles.) 

In agreement with Eq.~(\ref{eq_Gbound}), the ratio $G(t)/\muA$ for all operational definitions drops rapidly below unity.
The dashed line indicates the asymptotic power-law slope $-1$ \cite{SXM12,XWP12}. This can be understood by noting
that at imposed $\tau$ the shear strain freely diffuses in the liquid limit, 
i.e. $\langle \delta \gamhat^2 \rangle \sim t$, which implies that $G \approx \Gmodgt \sim 1/t$.
For both \NVtT- and \NPtT-ensembles we compare our key definition $\Gmodgt$, Eq.~(\ref{eq_Gmodgt}), 
to the observable $\Gmodgg$, Eq.~(\ref{eq_Gmodgg}). In both cases we confirm $\Gmodgt(t) \approx \Gmodgg(t)$ 
for larger times as suggested by the thermodynamic relation Eq.~(\ref{eq_dgamdtau}) which has been directly checked 
\cite{foot_dgamdtau}.

We have also computed the dimensionless correlation coefficient $\cmodgt(t)$, Eq.~(\ref{eq_cmodgt}), 
which characterize the correlation of the measured instantaneous shear strains $\gamhat$ and shear stresses $\tauhat$. 
As indicated for the \NPtT-ensemble (stars), $\cmodgt$ vanishes rapidly with $t$, i.e.  $\gamhat$ and $\tauhat$ are decorrelated.
The dash-dotted line indicates the power-law exponent $-1/2$ expected from Eq.~(\ref{eq_cmodgt_muA})
and $\Gmodgt(t) \sim 1/t$. 
Note that by plotting $\cmodgt(t)$ {\em vs.} $\Gmodgt(t)/\muA$, Eq.~(\ref{eq_cmodgt_muA}) has been directly
verified for both \NVtT- and \NPtT-ensembles (not shown).

The stress fluctuation formula, Eq.~(\ref{eq_Gmodtt}), is represented in Fig.~\ref{fig_Gt_Thigh} by squares and diamonds
for, respectively, \NVgT- and \NPgT-ensembles.
The bare shear modulus $\Gmodttbare$ {\em without} the impulsive correction $\muBcut$ for the
Born coefficient $\muB$ discussed in Sec.~\ref{theo_trunc} is given by small filled symbols.
The impulsive correction $\muBcut \approx 0.2$ (Table~\ref{tab_pLJ}) computed independently from the 
weighted radial pair correlation function, Eq.~(\ref{eq_histoLame}), is shown by a horizontal line. 
The corrected moduli $\Gmodtt=\Gmodttbare-\muBcut$ vanish indeed as expected.
See Ref.~\cite{XWP12} for a systematic numerical characterization of $\muBcut$ as a function of the reduced cutoff $\scut$.
As may also be seen there, a truncation correction is also necessary for the KA model.
The $\muBcut$ for different temperatures may be found in Table~\ref{tab_KA} for the KA model and in Table~\ref{tab_pLJ} 
for the pLJ model. We assume from now on that the impulsive corrections are properly taken into account without stating 
this technical point explicitly.

\subsection{Dynamical matrix harmonic network}
\label{res_dynmat}

\paragraph*{Introduction.}
We turn now to the opposite low-$T$ limit focusing specifically on the pLJ model at $T=0.001$.
Obviously, such deeply quenched colloidal glasses must behave as amorphous solids with a constant 
shear modulus for large times (albeit not infinite times) as sketched in Fig.~\ref{fig_sketch}(b)
by the top curve. 
Before presenting in Sec.~\ref{res_Tlow} the elastic properties of these glasses, we discuss first 
conceptually simpler substitute systems formed by {\em permanent} spring networks.
We do this for illustration purposes since questions related to ergodicity and ageing are by construction 
irrelevant and the thermodynamical relations reminded in Sec.~\ref{sec_theo} should hold rigorously.

\paragraph*{Network Hamiltonian.}
Assuming ideal harmonic springs the interaction energy reads
\begin{equation}
E = \sum_l \frac{1}{2} K_l \left(r_l - R_l\right)^2
\label{eq_harmonicnet}
\end{equation}
with $K_l$ being the spring constant and $R_l$ the reference length of spring $l$.
The sum runs over all springs $l$ between topologically connected vertices $i$ and $j$ of the network. 
We assume here a strongly and homogeneously connected network where every vertex $i$ is in contact with many neighbors $j$.
Such a network may be constructed from a pLJ configuration at $T=0.001$ keeping the particle positions for the positions 
of the vertices and replacing each LJ interaction $l$ by a spring of spring constant $K_l$ and reference length $R_l$.
The simplest choice we have investigated is to set $K_l \equiv 1$ and $R_l \equiv r_l$ which corresponds to a well-defined 
ground state of energy $E=0$ and pressure $P=0$ at $T=0$ (not shown). 
We discuss here instead a slightly more realistic choice of $K_l$ and $R_l$ which corresponds to the
{\em same} ``dynamical matrix" as the pLJ configuration at $T \to 0$ and $P=2$, i.e. the same second derivative of 
the total interaction potential with respect to the particle positions \cite{WTBL02,TWLB02}.
This choice imposes the setting 
\begin{eqnarray}
K_l & \equiv & \left.u^{\prime\prime}(r)\right|_{r=r_l}, \nonumber \\
R_l & \equiv & r_l - \left.\frac{u^{\prime}(r)}{u^{\prime\prime}(r)}\right|_{r=r_l}
\label{eq_KlRldynmat}
\end{eqnarray}
with $u(r)=u(s)$ being the interaction potential for reduced distances $\sijl = \rijl/\sigmaij \le \scut$. 
(Impulsive corrections at $\sijl = \scut$ are neglected here.) 

\paragraph*{Sampling.}
Taking advantage of the fact that all interacting vertices are known, these networks can be computed 
using global MC moves with longitudinal and transverse planar waves as described in Sec.~\ref{sec_algo}.
No local MC jumps for individual vertices have been added.
Note that although some $K_l$ and $R_l$ may even be negative, this does not affect the global or local
stability of the network. Presumably due to the fact that our spring constants are rather large,
we have not observed either any buckling instability if volume fluctuations are allowed.
For networks at the same pressure $P$, shear stress $\tau$ and temperature $T$ as the original pLJ configuration,
we obtain, not surprisingly, the same affine shear elasticity $\muA = 33.9$ 
as for the reference. The same applies for other simple averages.

\begin{figure}[t]
\centerline{\resizebox{1.0\columnwidth}{!}{\includegraphics*{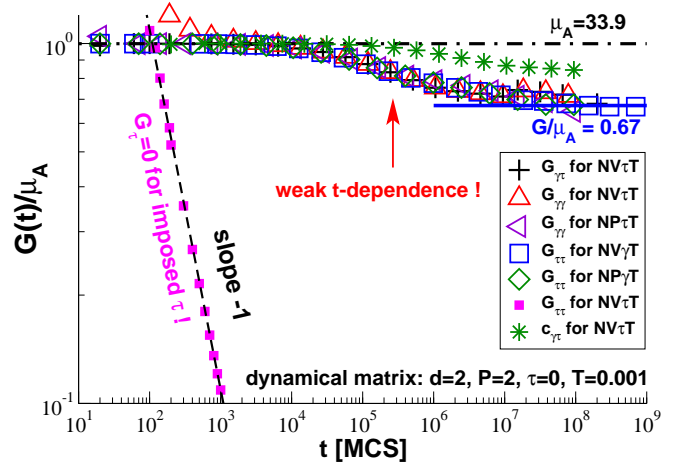}}}
\caption{Rescaled shear modulus $G(t)/\muA$ {\em vs.} sampling time $t$ for systems of 
permanent linear springs corresponding to the dynamical matrix of pLJ systems at 
$P=2$, $\tau=0$ and $T=0.001$.
As indicated by the bold line, we obtain $G/\muA \approx 0.67$ in the long-time limit.
In agreement with Eq.~(\ref{eq_muAmuFtau}) $\Gmodtt$ vanishes with time if it is computed 
in an ensemble where $\gamma$ fluctuates as shown for the \NVtT-ensemble (filled squares).
\label{fig_Gt_dynmat}
}
\end{figure}

\paragraph*{Shear modulus.}
Figure~\ref{fig_Gt_dynmat} presents various measurements of the shear modulus $G$ for different ensembles
as a function of sampling time $t$. As in Sec.~\ref{res_Thigh}, we use gliding averages 
and the vertical axis has been rescaled by the affine shear elasticity $\muA$.
As can be seen, we obtain in the long-time limit $G/\muA \approx 0.67$ (bold line) irrespective of whether
the shear modulus is determined from $\Gmodgt$ or $\Gmodgg$ in the \NVtT- or \NPtT-ensembles
or using $\Gmodtt$ in the \NVgT- or \NPgT-ensembles at fixed shear strain $\gamma$.
Note that even in these cases the shear modulus decreases first with time albeit the
network is perfectly at thermodynamic equilibrium and no ageing occurs by construction.
As shown by the dash-dotted line, $G(t)/\muA \to 1$ for short times,
i.e. in this limit the response is affine. 
As indicated by the stars, we have also computed the correlation coefficient $\cmodgt(t)$ as a function of time.
From $\cmodgt =1$ for small $t$, it decreases somewhat becoming 
$\cmodgt \approx \sqrt{0.67} \approx 0.82$ for large times in agreement with Eq.~(\ref{eq_cmodgt_muA}).
If $\Gmodtt$ is computed in the ``wrong" \NVtT- or \NPtT-ensembles, 
it is seen to vanish rapidly with time as shown by the filled squares. This decay is of 
course expected from Eq.~(\ref{eq_muAmuFtau}) as discussed in Sec.~\ref{theo_stressfluctu} 
and Sec.~\ref{theo_Legendre}.

\begin{figure}[t]
\centerline{\resizebox{1.0\columnwidth}{!}{\includegraphics*{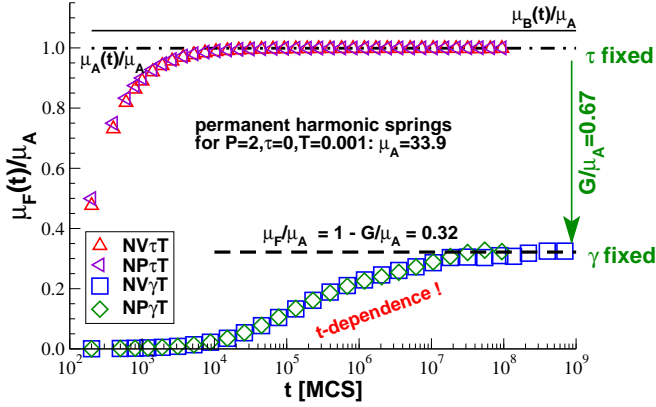}}}
\caption{Shear stress fluctuations $\muF(t)/\muA$ with $\muA = 33.9$ for the same systems as in Fig.~\ref{fig_Gt_dynmat}.
We observe $\muF(t)/\muA \to 1$ (dash-dotted line) for imposed $\tau$,
while $\muF(t)/\muA \to 1 - G/\muA \approx 0.32$ (dashed line) for imposed $\gamma$.
Note the broad time-dependence of $\muF(t)$ in the latter cases.
\label{fig_muFt_dynmat}
}
\end{figure}

\paragraph*{Shear-stress fluctuations.}
To emphasize the latter point we show in Fig.~\ref{fig_muFt_dynmat} the stress fluctuations $\muF(t)$
for different ensembles. This is also done to have a closer look at the time dependence of various
contributions to $\Gmodtt(t)$. As before, the vertical axis has been rescaled with $\muA \equiv 33.9$.
We also indicate the rescaled values of $\muB(t)$ (thin top line) and $\muA(t) = \muB(t) - \Pex(t)$ (dash-dotted line)
computed by a gliding average over time intervals $[t_0,t_1=t_0+t]$. Being simple averages both properties do not
depend on the ensemble probed (not shown) and become virtually immediately $t$-independent as can be seen from 
the indicated horizontal lines.
This is quite different for the different shear stress fluctuations $\muF$ presented which increase monotonously
from $\muF \approx 0$ for short times to their plateau value for long times. In agreement with Eq.~(\ref{eq_muAmuFtau}) 
we find that $\left.\muF(t)\right|_{\tau} \to \muA$ for \NVtT- and \NPtT-ensembles and as predicted by 
Eq.~(\ref{eq_muFtransform}) we confirm $\left.\muF(t)\right|_{\gamma} \to \muA - G$ for \NVgT- and \NPgT-ensembles.
The asymptotic limit is reached after about $10^4$ MCS in the first case for imposed $\tau$, 
but only after $10^7$ for imposed $\gamma$. This is due to the fact that $\left.\muF\right|_{\tau}$ 
corresponds to the fluctuations of the self-contribution of individual springs as discussed in Sec.~\ref{theo_Legendre}, 
while $\left.\muF\right|_{\gamma}$ also contains contributions from stress correlations of different springs. 
It should be rewarding to analyze in the future finite system-size effects and the time dependence 
of three-dimensional (3D) systems using the dynamical matrix associated to the KA model in the $T \to 0$ limit.

\begin{figure}[t]
\centerline{\resizebox{1.0\columnwidth}{!}{\includegraphics*{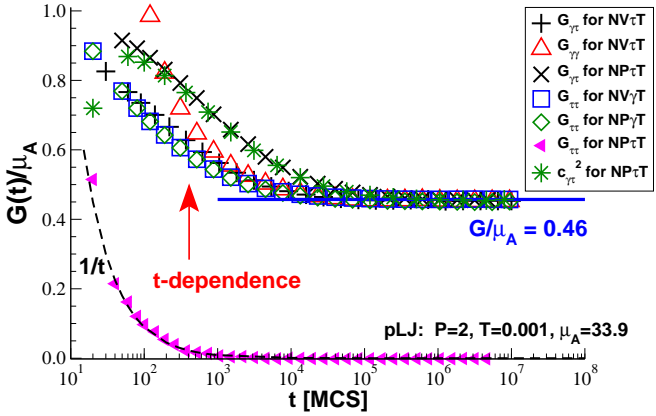}}}
\caption{$\Gmodgt(t)/\muA$, $\Gmodgg(t)/\muA$ and $\Gmodtt(t)/\muA$ for pLJ beads
in the low-temperature limit for $P=2$, $\tau=0$ and $T =0.001$ where $\muA \approx 33.9$. 
Qualitatively, the data compares well with the various shear moduli presented in Fig.~\ref{fig_Gt_dynmat}. 
The filled triangles show $\Gmodtt$ for the \NPtT-ensemble which is seen to vanish rapidly with sampling time $t$.
The dashed line indicates a $1/t$-decay.
\label{fig_Gt_Tlow}
}
\end{figure}

\subsection{Low-temperature glass limit}
\label{res_Tlow}

We return now to the colloidal glasses where the topology of the interactions is, of course,
{\em not} permanently fixed (at least not at finite temperatures) and the shear stresses may 
thus fluctuate more strongly.
As an example, we present in Fig.~\ref{fig_Gt_Tlow} the shear modulus $G(t)$ as a function of
sampling time $t$ for pLJ beads at mean pressure $P=2$, mean shear stress $\tau= 0$ and
mean temperature $T=0.001$, i.e. the same conditions as in the previous subsection.
All simple averages, such as the affine shear elasticity $\muA=33.9$, are the same for all 
ensembles albeit different quench protocols have been used.
For all examples given we use local MC jumps with $\drmax=0.01$ which allows to keep
a small, but still reasonable acceptance rate $A \approx 0.161$. We have also performed runs with
additional global MC moves using longitudinal and transverse plane waves. 
This yields similar data which is shifted horizontally to the left (not shown).

As one would expect, the shear moduli presented in Fig.~\ref{fig_Gt_Tlow} are similar 
to the ones shown in Fig.~\ref{fig_Gt_dynmat} for the permanent spring network. 
The filled triangles show $\Gmodtt$ computed in the ``wrong" \NPtT-ensemble.
As expected from Eq.~(\ref{eq_muFtransform}) these values vanish rapidly with time.
If on the other hand $\Gmodgt$, $\Gmodgg$ and $\Gmodtt$ are computed in their natural ensembles,
all these measures of $G$ are similar and approach, as one expects for such a low temperature,
the same finite asymptotic value $G/\muA \approx 0.46$ indicated by the bold horizontal line. 
The stars indicate the squared correlation coefficient $\cmodgt^2(t)$ which is seen to 
collapse perfectly on the rescaled modulus $\Gmodgt(t)/\muA$ which shows that 
Eq.~(\ref{eq_cmodgt_muA}) even holds for small times before the thermodynamic plateau has been reached.
Similar reduced shear moduli $\Gmodtt/\muA$ have been also obtained for the KA model in the low-$T$ limit 
as may be seen from Table~\ref{tab_KA}. 
These results confirm that about half of the affine shear elasticity is released by the shear stress fluctuations
in agreement with Refs.~\cite{WTBL02,TWLB02}. 
Interestingly, this value is slightly smaller as the corresponding value for the permanent 
dynamical matrix model presented above.  
Since some, albeit very small, rearrangement of the interactions must occur for the pLJ particles 
at finite temperature, this is a reasonable finding. 
%

\begin{figure}[t]
\centerline{\resizebox{1.0\columnwidth}{!}{\includegraphics*{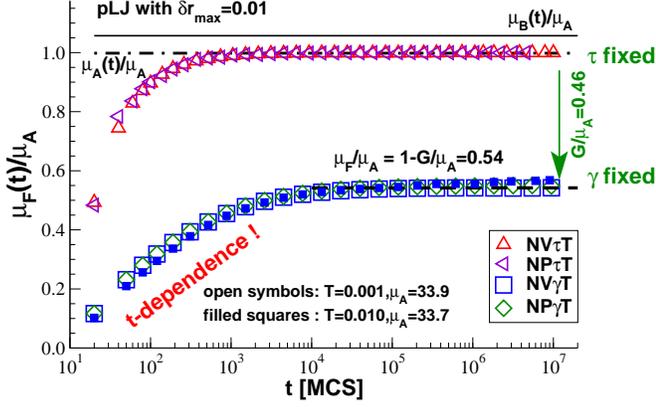}}}
\caption{Reduced shear stress fluctuation $\muF(t)/\muA$ for pLJ beads using the same 
representation as in Fig.~\ref{fig_muFt_dynmat}. The small filled squares refer to data
obtained for $T=0.010$ using the \NVgT-ensemble, all other data to $T=0.001$.
The reduced Born coefficient $\muB(t)/\muA$ (thin line) and the affine shear elasticity 
$\muA(t)/\muA$ (dash-dotted line) are essentially time independent 
while $\muF(t)/\muA$ approaches its large-$t$ limit (dashed line) from below.
\label{fig_muFt_Tlow}
}
\end{figure}

The rescaled shear stress fluctuations $\muF(t)/\muA$ for the same ensembles are presented in Fig.~\ref{fig_muFt_Tlow}.
While the two-point correlations $\muB(t)/\muA$ (thin horizontal line) and $\muA(t)/\muA$ (dash-dotted horizontal line) 
do not dependent on time, one observes again that $\muF(t)/\muA$ increases monotonously from zero to its
long-time asymptotic plateau value. In agreement with Eq.~(\ref{eq_muAmuFtau}) we find $\left.\muF(t)\right|_{\tau} \to \muA$ 
for the \NVtT \ and the \NPtT \ ensembles, while $\left.\muF(t)\right|_{\gamma} \to \muA - G$ (dashed horizontal line) for 
the \NVgT \ and the \NPgT \ ensembles.
(Similar behavior has been obtained for the KA model in the \NVgT-ensemble for temperatures $T \le 0.1$.)
As may be better seen for the data set obtained for $T=0.010$ using the \NVgT-ensemble (small filled squares)
where $\muA \approx 33.7$, the stress fluctuations $\left.\muF(t)\right|_{\gamma}$ do not rigorously become constant, 
but increase extremely slowly on the logarithmic time scales presented. 
There is, hence, even at low temperatures always some sampling time dependence
if the interactions are not permanently fixed as for the topologically fixed networks discussed above. 
However, this effect becomes only sizeable above $T=0.2$ for the KA model 
and above $T=0.1$ for the pLJ model.
We shall now attempt to characterize this temperature dependence.

\subsection{Scaling with temperature}
\label{res_Tvari}

\begin{figure}[t]
\centerline{\resizebox{1.0\columnwidth}{!}{\includegraphics*{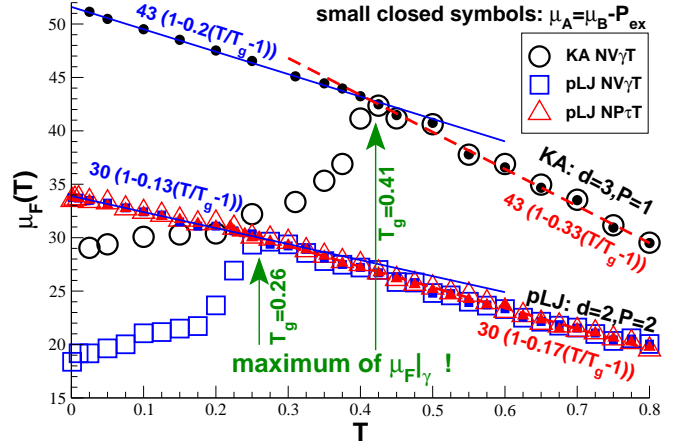}}}
\caption{Stress fluctuation $\muF$ (open symbols) and affine shear elasticity $\muA$ (small filled symbols) 
{\em vs.} temperature $T$ for both models. The indicated values have been obtained from the longest 
simulation runs available for a given temperature. $\muA$ decays roughly linearly with $T$ as shown by 
bold solid lines for the low-$T$ and by dashed line for the high-$T$ regime.  
For constant-$\tau$ systems and for all systems with $T \ge \Tglass$ we confirm that $\muF \approx \muA$, 
while $\muFgam$ (open spheres and squares) is seen to be non-monotonous with a maximum at $\Tglass$.
\label{fig_muAmuF_T}
}
\end{figure}

\paragraph*{Stress fluctuations.}
We turn now to the characterization of the temperature dependence of the shear modulus $G$ and various 
related properties. We focus first on the best values obtained for the longest simulation runs available.
Figure~\ref{fig_muAmuF_T} presents the stress fluctuations obtained for the KA model using the \NVgT-ensemble (open spheres) 
and for the pLJ model using \NVgT- and the \NPtT-ensembles. 
The affine shear elasticity $\muA$ obtained for each system is represented by small filled symbols.
As shown for both models by a bold solid line for the low-$T$ and by a dashed line for the high-$T$ regime,
$\muA(T)$ decays roughly linearly with $T$. Interestingly, the slopes match precisely at $T=\Tglass$ for both models.
As may be seen from the large triangles for the \NPtT-ensemble, we confirm for all temperatures
that $\muFtau \approx \muA$ as expected from Eq.~(\ref{eq_muAmuFtau}).
For systems without box shape fluctuations, it is seen that $\muFgam$ becomes {\em non-monotonous}
with a clear maximum at $T \approx \Tglass$. In the liquid regime for $T > \Tglass$ where the boundary conditions
do not matter, we obtain $\muFtau \approx \muFgam \approx \muA$, i.e. all shear stress fluctuations decay with temperature.
Below $\Tglass$ the boundary constraint ($\gamma=0$) becomes more and more relevant with decreasing $T$
which reduces the shear stress fluctuations. 
According to Eq.~(\ref{eq_muFtransform}), $\muFgam$ must thus decrease with increasing $G$ 
as the system is further cooled down. 
That $\muFgam$ must necessarily be non-monotonous with a maximum at $\Tglass$ is one of the 
central results of the presented work.

\begin{figure}[t]
\centerline{\resizebox{1.0\columnwidth}{!}{\includegraphics*{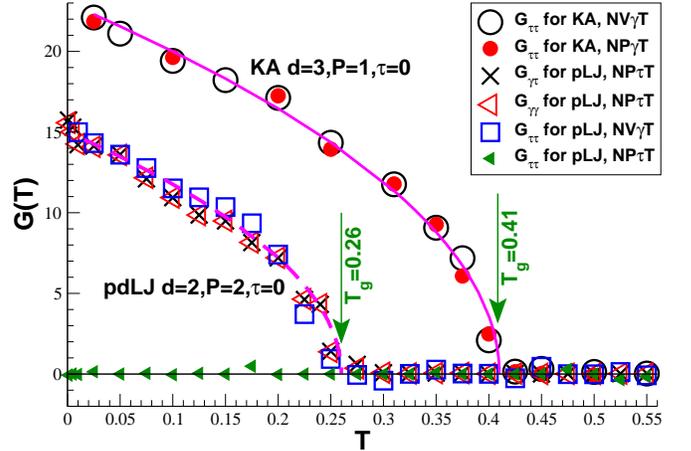}}}
\caption{Unscaled shear modulus $G$ as a function of temperature $T$ for both models.
For the pLJ model the values $\Gmodtt$ obtained in the \NVgT \ ensemble (squares)
are essentially identical to the moduli $\Gmodgt$ (crosses) and $\Gmodgg$ (triangles)
obtained for constant $\tau$.
As shown by the small filled triangles, $\Gmodtt$ vanishes in the \NPtT-ensemble for all $T$
in agreement with Eq.~(\ref{eq_muAmuFtau}). 
The solid and the dashed line indicate the cusp-singularity, Eq.~(\ref{eq_cuspcont}),
for respectively the KA model and the pLJ beads
\cite{Mezard10,Yoshino12}. 
\label{fig_GT}
}
\end{figure}

\paragraph*{Shear modulus as a function of temperature.}
The temperature dependence of the (unscaled) shear moduli $G(T)$ is presented in Fig.~\ref{fig_GT}.
Data obtained by MD simulation of the KA model ($P \approx 1$, $\Tglass \approx 0.41$) are indicated
for the \NVgT-ensemble (open spheres) and the \NPgT-ensemble (filled spheres).
All other results refer to the best values obtained by MC 
simulations of the pLJ model ($P=2$, $\Tglass \approx 0.26$) using both local and global moves and
different ensembles as indicated.
For the pLJ model the values $\Gmodtt$ obtained in the \NVgT \ ensemble (squares)
are seen to be essentially identical for all $T$ to the moduli $\Gmodgt$ (crosses) and $\Gmodgg$ (triangles)
obtained in ensembles with strain fluctuations.
As shown by the small filled triangles, $\Gmodtt$ vanishes in the \NPtT \ ensemble for all $T$
in agreement with Eq.~(\ref{eq_muAmuFtau}). 
(We remind that the impulsive correction $\muBcut$ has to be taken into account if $\Gmodtt$ is used.)
Decreasing the temperature further below $\Tglass$ the shear moduli are seen 
to increase rapidly for both models.
As shown by the solid and the dashed lines both models are well fitted
by a cusp-singularity 
\begin{equation}
G(T) \approx g_1 \left(1-T/\Tglass \right)^{1/2} + g_0 \mbox{ for } T < \Tglass 
\label{eq_cuspcont}
\end{equation}
with fit constants $g_1 \approx 23$ for the KA model and $g_1 \approx 15$ for the pLJ model 
where we have set in both cases $g_0=0$.
Confirming the MD simulations by Barrat {\em et al.} \cite{Barrat88}, this suggests that the transition is very sharp, 
albeit {\em continuous} in qualitative agreement with the predictions from Ref.~\cite{Mezard10,Yoshino12}.
If we fit the data close to $\Tglass$ with an additional off-set $g_0$, a slightly {\em negative} value is obtained 
which is not compatible with MCT  \cite{GoetzeBook,Klix12}. However, admittedly both the number of data points close 
to $\Tglass$ and their precision (due to the small number of independent configurations) 
are yet not sufficient to rule out completely a positive, albeit (presumably) small off-set. 

\begin{figure}[t]
\centerline{\resizebox{0.95\columnwidth}{!}{\includegraphics*{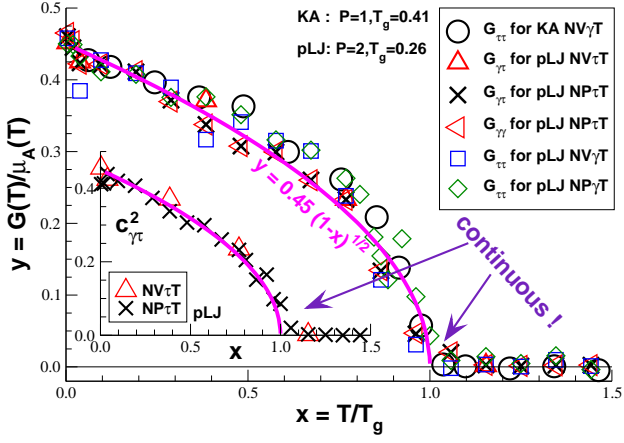}}}
\caption{Rescaled shear modulus $y = G/\muA$ {\em vs.} reduced temperature $x = T/\Tglass$ 
for the KA model at $P=1$ (\NVgT-ensemble) and the pLJ model at $P=2$ (\NVtT-, \NPtT-, \NVgT- and \NPgT-ensembles).
The bold line indicates again a {\em continuous} cusp-singularity. 
Inset: Squared correlation coefficient $\cmodgt^2$ for \NVtT- and \NPtT-ensembles.
\label{fig_GT_rescal}
}
\end{figure}

A slightly different representation of the data is given in Fig.~\ref{fig_GT_rescal} where the 
reduced shear modulus $y=G(T)/\muA(T)$ is plotted as a function of the reduced temperature 
$x= T/\Tglass$ for both models. Using the independently determined glass transition temperature $\Tglass$ 
and affine shear elasticity $\muA(T)$ as scales to make the axes dimensionless,
the rescaled data are found to collapse ! This is of course a remarkable and rather unexpected result 
considering that two different models in two different dimensions have been compared. 
The bold line indicates the cusp-singularity $y = 0.45 (1-x)^{1/2}$ with a prefactor which is compatible 
to the ones used in Eq.~(\ref{eq_cuspcont}) considering the typical values of $\muA(T)$ in both models.
Whether this striking collapse is just due to some lucky coincidence or makes manifest a more general 
universal scaling is impossible to decide at present.
(Please note that the corresponding data for the compression modulus shown in Fig.~\ref{fig_KT} does not scale.)
No rescaling of the vertical axis is necessary for the squared correlation coefficient $\cmodgt^2$ shown in the inset
for two different ensembles of the pLJ model.  The same continuous cusp-like decay (bold line) is found as for the modulus $G$. 
This is again consistent with the thermodynamic reasoning put forward at the end of Sec.~\ref{theo_strainfluctu}, Eq.~(\ref{eq_cmodgt_muA}).

\begin{figure}[t]
\centerline{\resizebox{1.0\columnwidth}{!}{\includegraphics*{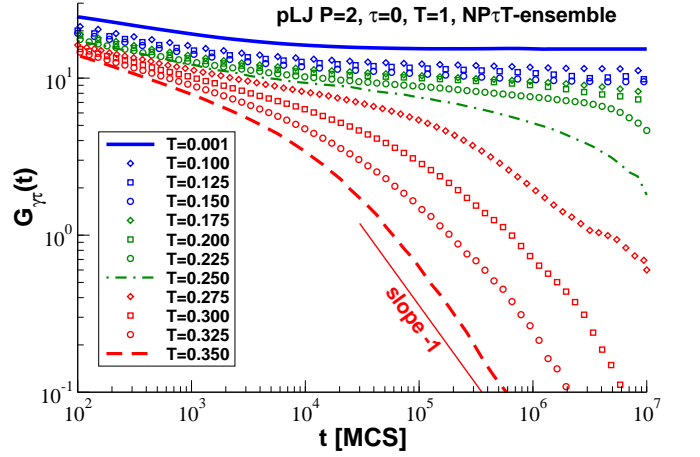}}}
\caption{Unscaled shear modulus $\Gmodgt$ as a function of sampling time $t$ in MCS obtained for the pLJ model in the \NPtT-ensemble
for different temperatures as indicated. The thin line indicates the power-law slope $-1$ for large times in the liquid regime.
A clear shoulder (on the logarithmic scales used) can only be seen below $T \approx 0.2$
and only below $T=0.01$ one observes a $t$-independent plateau over two orders of magnitude.
We emphasize that we present here a sampling time effect and {\em not} a time correlation function. 
The same sampling time effect is observed after additional tempering.
\label{fig_Gt_T}
}
\end{figure}

\paragraph*{Sampling time dependence.}
A word of caution must be placed here.
The results presented in Figs.~\ref{fig_muAmuF_T}-\ref{fig_GT_rescal} correspond to the longest 
simulation runs we have at present been able to perform. Obviously, this does {\em not} mean 
that they correspond to the limit for asymptotically long sampling times, if the latter limit exists, 
or at least an intermediate plateau of respectable period of time. 
Focusing on the pLJ model in the \NPtT-ensemble we attempt in Fig.~\ref{fig_Gt_T} to give a tentative characterization 
of the sampling time dependence. We plot $\Gmodgt(t)$ as a function of $t$ for different temperatures as indicated. 
To make the data comparable all simulations for $T \ge 0.1$ have been performed with local MC jumps of same 
maximum distance $\drmax=0.1$. 
As we have already seen in Fig.~\ref{fig_Gt_Thigh}, the shear modulus decays inversely with time
in the liquid regime at high temperatures. This limit is indicated by the thin line.
With decreasing $T$ the (unscaled) shear modulus increases and a shoulder develops.
Note that even for $T=0.25$, i.e. slightly below the glass transition temperature $\Tglass =0.26$ estimated via dilatometry 
(Fig.~\ref{fig_quench_rho}), $\Gmodgt(t)$ decays strongly with time. 
A more or less $t$-independent shoulder (on the logarithmic scales used) can only be seen below $T \approx 0.2$. 
We emphasize that we present here a sampling time effect expressing the fact that the configuration space is more and more 
explored with increasing $t$ and {\em not} a time correlation function \cite{AllenTildesleyBook}. 
Please note also that this time dependence has nothing to do with equilibration problems or ageing effects. 
Time translational invariance is perfectly obeyed in our simulations as we have checked by rerunning the sampling 
{\em after} additional tempering over at least $10^7$ MCS for all temperatures. 
Similar $t$-dependencies for different temperatures $T$ have also been observed for $\Gmodgt(t)$ and $\Gmodtt(t)$ for the pLJ model
and for $\Gmodtt(t)$ for the KA model (not shown). 

\begin{figure}[t]
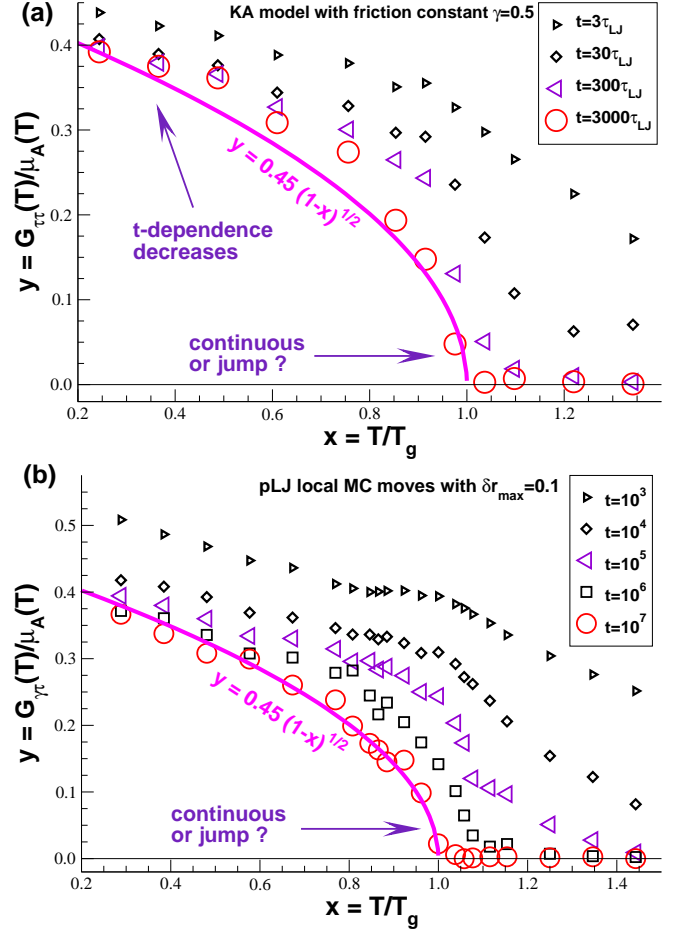

\centerline{\resizebox{1.0\columnwidth}{!}{\includegraphics*{fig13a}}}
\centerline{\resizebox{1.0\columnwidth}{!}{\includegraphics*{fig13b}}}
\caption{Reduced shear modulus $G(T)/\muA(T)$ as a function of reduced temperature $x = T/\Tglass$ for different sampling times $t$ as indicated:
{\bf (a)} $\Gmodtt(T)$ for the KA model (\NVgT-ensemble), {\bf (b)} $\Gmodgt(T)$ for the pLJ model (\NPtT-ensemble).
For both models the glass transition is seen to become sharper with increasing $t$.
The continuous cusp-singularity is indicated by the bold line.
More points in the vicinity of $x \approx 1$ and large sampling times are needed in the future to decide 
whether the shear modulus is continuous or discontinuous at the glass transition.
\label{fig_GT_t}
}
\end{figure}

A different representation of the data for both models is shown in Fig.~\ref{fig_GT_t} where the reduced shear modulus 
$y =G(T)/\muA(T)$ is plotted against the reduced temperature $x = T/\Tglass$ for different sampling times $t$ as indicated. 
Similar behavior is found for both models. The shear modulus decreases with $t$ and this the more the larger the temperature $T$.
For smaller temperatures the different data sets appear to approach more readily as one expects from the $t$-independent 
shoulder for the pLJ model in Fig.~\ref{fig_Gt_T}.
As a consequence the glass transition is seen to become sharper with increasing $t$.
The data approach the continuous cusp-singularity indicated by the bold line.
However, it is not clear from the current data if the data converge to an at least {\em intermediately} stable behavior. 
Longer simulation runs are clearly necessary for both models to clarify this issue.
Similar data have also been obtained for the pLJ model from the independent determination of $\Gmodgt(t)$ and $\Gmodgg(t)$ 
in the \NVtT-ensemble and for $\Gmodtt(t)$ in the \NVgT- and \NPgT-ensemble.

\section{Conclusion}
\label{sec_conc}
\subsection{Summary}
\label{conc_summary}
The stress fluctuation formalism is a powerful method for computing elastic moduli for canonical ensembles
using simulation boxes of constant volume and shape \cite{Hoover69,Lutsko89,FrenkelSmitBook,SBM11}. 
Focusing on the shear modulus $G$ of two  well-known glass-forming colloidal model systems in $d=3$ and $d=2$ dimensions, 
which we have sampled by means of MD and MC simulations, we have addressed the general question of whether the stress fluctuation method may be used 
through a solid-liquid transition where a well-defined displacement field ceases to be defined.
Deliberately setting the experimentally motivated linear regression Eq.~(\ref{eq_Gmodgt}) as the fundamental definition, 
the shear modulus $G$ has been computed from the strain and stress fluctuations 
(assuming a fixed finite measurement time window $\tmeas$)
in ensembles where either a shear strain $\gamma=0$ (\NVgT- and \NPgT-ensembles) 
or a (mean) shear stress $\tau=0$ (\NVtT- and \NPtT-ensembles) have been imposed (Fig.~\ref{fig_sketch}). 
Working at constant mean normal pressure we have computed and compared the temperature 
dependence for various simple averages and fluctuations contributing to the shear modulus in different ensembles.
For the KA model only the currently available results for \NVgT- and \NPgT-ensembles have been presented.

As has been stressed in Sec.~\ref{theo_stressfluctu} and Sec.~\ref{theo_Legendre}, the stress fluctuation representation $\Gmodtt$ 
at fixed shear strain does not rely {\em in principle} on a solid-like reference state for a displacement field of tagged particles
if thermodynamics can be assumed to hold. 
It thus holds formally through the glass-transition temperature $\Tglass$ up to the liquid regime 
(albeit with a trivial value $G=0$) as does the better known relation for the compression modulus $K$ 
(Appendix~\ref{app_K}) \cite{RowlinsonBook}. 
As emphasized in Sec.~\ref{theo_trunc}, impulsive corrections due to the truncation of the pair potentials 
have to be properly taken into account, especially at high temperatures (Fig.~\ref{fig_Gt_Thigh}), 
for the precise determination of the affine shear elasticity $\muA = \muB -\Pex$ \cite{XWP12}.
Confirming Eq.~(\ref{eq_Gmods}), it has been shown for the pLJ model that $\Gmodtt$ at $\gamma=0$ 
has the {\em same} long-time behavior as the conceptually more direct observables $\Gmodgt$ \ and $\Gmodgg$ \ 
obtained using the strain fluctuations at constant mean shear stress $\tau=0$.
The standard thermodynamic relations comparing different ensembles appear thus to hold also {\em in practice} 
(at least within our numerical precision) through the glass transition for all $T$ and this albeit  
\begin{enumerate}
\item
some degrees of freedom get quenched at low temperatures on the time window $\tmeas$ computationally accessible,
\item
 all operational definitions of $G$ are transient in the sense that they vanish for $\tmeas \to \infty$
(Figs.~\ref{fig_Gt_T} and \ref{fig_GT_t})
and 
\item
it is not self-evident that in the latter limit and for large temperatures
$\gamma$ and $\tau$ can still be treated as a pair of conjugated thermostatistical variables.
\end{enumerate}
The latter point compactly epitomized by the thermodynamic relation Eq.~(\ref{eq_dgamdtau}) has been shown to 
hold, however, with remarkable precision up to very high temperatures for the pLJ model.

As predicted by general Legendre transformation, $\Gmodtt$ vanishes for all temperatures $T$, if $\tau$ rather than $\gamma$ 
is imposed (Figs.~\ref{fig_Gt_dynmat},~\ref{fig_Gt_Tlow} and \ref{fig_GT}). 
The shear-stress fluctuations $\left.\muF\right|_{\tau}$ are thus given by the affine response under an external load $\muA$
(Fig.~\ref{fig_muAmuF_T}), 
i.e. $\left.\muF\right|_{\tau}$ reduces to a simple two-point pair correlation function if pair potentials are considered
as discussed in Sec.~\ref{theo_Legendre}.
The same holds above $\Tglass$ for constant $\gamma$, since the boundary conditions are irrelevant for the liquid state.
This symmetry with respect to the boundary conditions ($\tau \leftrightarrow \gamma$) is broken at the glass transition
for a large, but finite $\tmeas$:
at constant $\gamma$ the stress fluctuations reveal a strong non-monotonous behavior with a clear maximum at $\Tglass$ 
(Fig.~\ref{fig_muAmuF_T}). 
The shear modulus is the order parameter characterizing this symmetry breaking.
Alternatively, as we have discussed in Sec.~\ref{theo_Legendre}, $\Gmodtt/\muA \equiv 1 - \left.\muF\right|_{\gamma}/\muA$ 
may be seen as an order parameter comparing the ratio of the non-affine to the affine shear responses.
Since $\muA > \left.\muF\right|_{\gamma}$ for $T < \Tglass$, this implies that the stress fluctuations contain 
higher-order correlations {\em reducing} the free energy of the system.

The increase of $G$ below $\Tglass$ is reasonably fitted for both models by a {\em continuous} cusp singularity,
Eq.~(\ref{eq_cuspcont}), in qualitative agreement with recent replica calculations \cite{Mezard10,Yoshino12}. 
A jump discontinuity, as suggested by mode-coupling theory \cite{GoetzeBook,Klix12} and another replica theory \cite{Szamel11}, 
is not compatible with the currently available data shown in Fig.~\ref{fig_GT} and Fig.~\ref{fig_GT_rescal}.
However, as shown in Fig.~\ref{fig_Gt_T} and  Fig.~\ref{fig_GT_t}, our data depend strongly on sampling time $t$ and the computation of
larger $t$ could lead to a sharper transition. Thus a jump discontinuity cannot be ruled out completely.
(We are not aware of any other investigation of the sampling time effect for the shear modulus.)
At present we believe, however, that it would be surprising if our data could be reconciled with a discontinuity at $\Tglass$.
It seems more likely that below $\Tglass$, where the choice of the boundary conditions does matter as shown, 
the definition of the glassy shear modulus used in Refs.~\cite{GoetzeBook,Klix12,Szamel11} do not correspond to 
the key operational definition $\Gmodgt$ of the shear modulus considered by us. 
In any case, our work shows that any theory and numerical scheme used to determine the shear modulus using a stress or 
strain fluctuation relation (in the low-wavevector limit or at a small finite $q$) must cleary specify and take into account
whether the shear stress or the shear strain are macroscopically imposed. Otherwise, such an approach is void.
 
\subsection{Outlook}
\label{conc_outlook}

\paragraph*{Beyond the presented work.}
Future work should clearly focus on the more detailed description of the sampling time dependence.
By extrapolating appropriately for the large-$t$ asymptotic behavior, this may allow to settle the theoretical debate.
As emphasized above, one shortcoming of the present work is that the time-scale used in our MC simulations 
was slightly arbitrary, especially if additional non-local particle moves are used in the low-$T$ limit and 
if box volume and shape changing strains are included. The comparison of $t$-dependent properties for different
temperatures $T$ becomes thus delicate. 
Our MC study for the 2D soft particles should thus in any case be recomputed with Langevin thermostat MD 
dynamics as used for the data of the KA model presented. The $t$-dependence of the latter model, which is a
better glass former than the 2D pLJ model \cite{foot_dynamics}, 
has still to be worked out for the key operational definition $\Gmodgt$ in the \NVtT- or \NPtT-ensembles \cite{KA}.
We plan also in the near future to reconsider more carefully our previous investigation of the glass transition of polymer melts \cite{SBM11}
taking properly into account the impulsive truncation corrections and comparing the shear moduli $\Gmodgt\approx \Gmodgg$ (\NPtT-ensemble) 
and $\Gmodtt$ (\NPtT-ensemble) around $\Tglass$.

In addition, it should be rewarding to compare our shear moduli with the values obtained from the displacement 
of particles following the procedure chosen in Ref.~\cite{Klix12}. This procedure relies on the idea that
the particles fluctuate around a well-defined reference position. In the low-$T$ limit where the interaction
network can be replaced by the dynamical matrix, this should yield the same results. 
However, for larger temperatures where the harmonic approximation must break down,
this approach is questionable (both for the analysis of experimental and computational data)
and should be compared to the moduli obtained directly from the strain and stress fluctuations of the overall simulation box.

\begin{figure}[t]
\centerline{\resizebox{1.0\columnwidth}{!}{\includegraphics*{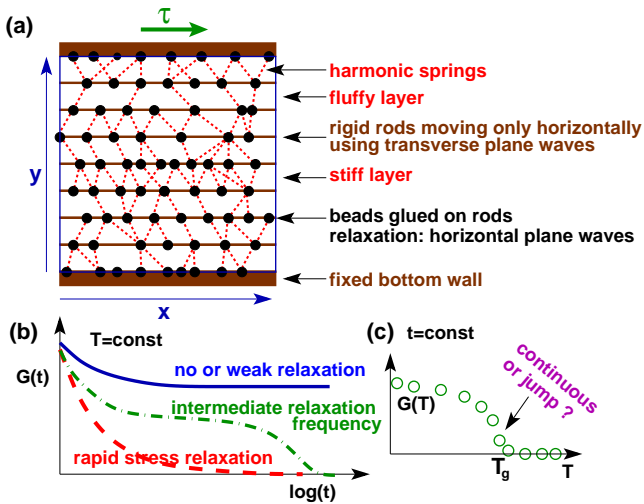}}}
\caption{Outlook: 
{\bf (a)}
A simplified scalar model for permanent and transient networks where beads (filled spheres) are glued 
transiently on rigid rods which are only allowed to move horizontally \cite{scalarGmodel}. 
{\bf (b)}
Shear modulus $G(t)$ as a function of sampling time $t$ at fixed temperature $T$ 
for three different mean barrier heights $\la \Ebarrier \ra$. 
{\bf (c)}
Possible shape of $G(T)$ {\em vs.} temperature $T$ for a fixed sampling time $t$.
\label{fig_band}
}
\end{figure}

\paragraph*{A simple scalar model.}
We are currently working on an extremely simplified scalar model for permanent and transient networks
in $d=2$ dimensions which may help to clarify the theoretical debate \cite{scalarGmodel}. 
As sketched in the panel (a) of Fig.~\ref{fig_band}, in this model beads (filled spheres) are glued on 
rigid rods which are only allowed to move horizontally. In its most simple implementation this is done 
by means of global MC moves using transverse plane waves with a wavevector pointing in the $y$-direction.  
The glued beads are connected by ideal harmonic springs where the spring constants $K_l$ and reference
lengths $R_l$ are chosen randomly according to fixed, narrow or broad distributions.
The stiffness of the contacts between rods may differ (even including negative $K_l$ and $R_l$ as for
the dynamical matrix studied in Sec.~\ref{res_dynmat}). 
The shear modulus may be computed either from the  shear strain fluctuations in the \NVtT- or 
the shear stress fluctuations in the \NVgT-ensemble. 
The shear stresses can be relaxed by either allowing the particles to move along the rods using 
longitudinal plane waves in the $x$-direction and/or by breaking and reconnection of springs.
The latter step is currently done by applying a chemical potential for the springs.
Physically, the study of such transient networks is motivated by systems of 
hyperbranched polymer chains with sticky end-groups \cite{Friedrich10} or 
telechelic polymers in water-oil emulsions \cite{Porte01b,Porte03}.
The associated relaxation frequency $\omega_l \propto \exp(-\Ebarrier/T)$ for a given spring $l$
is assumed to be set by an activation barrier $\Ebarrier$ which may differ for each spring.
Depending on the distribution of this activation barrier the shear modulus $G(t)$ thus decays more or less
rapidly with sampling time $t$ as sketched in panel (b) of Fig.~\ref{fig_band}.
The shear modulus becomes constant, if the random percolating network is permanent 
($\omega_l = 0$, $\Ebarrier/T \to \infty$) or the relaxation negligible (top solid line). 
The system behaves as a liquid (bottom line) for large (mean) $\la \omega_l \ra$, while a shoulder is seen 
in the intermediate frequency range. While this is confirmed by preliminary simulations,
we do at present not understand the role of the quenched noise injected in the model
which is seen to strongly influence the shear stress fluctuations $\muF$ 
(especially if fluffy layers happen to appear).
Also the understanding of finite system-size effects is of course crucial for such a low-dimensional model. 
(Only systems containing $N=10^4$ particles have been considered so far.)
As sketched in panel (c) the key question is to describe accurately for asymptotically long simulation 
runs and asymptotically large box sizes the shape of $G(T)$ around the glass transition temperature $\Tglass(\langle\Ebarrier\rangle)$ 
where the shear modulus vanishes and $\left.\muF\right|_{\gamma}$ should again have a maximum.

\begin{acknowledgments}
H.X. thanks the CNRS and the IRTG Soft Matter for supporting her sabbatical stay in Strasbourg,
P.P. the  R\'egion Alsace and the IRTG Soft Matter and F.W. the DAAD for funding.
We are indebted to A.~Blumen (Freiburg) and H.~Meyer, O.~Benzerara and J. Farago (all ICS, Strasbourg) for helpful discussions.
\end{acknowledgments}

\appendix
\section{Compression modulus}
\label{app_K}

\subsection{Operational definitions}
\label{app_Kstrainfluctu}

We have focused in the main part of this paper on the shear modulus $G$ of our glass-forming model 
systems and the comparison of the three observables $\Gmodgt$, $\Gmodgg$ and $\Gmodtt$.
Being isotropic the complete macroscopic elastic description of our systems only requires 
the numerical determination of {\em one} additional elastic modulus, the isothermal compression modulus 
\cite{LandauElasticity,Callen,ChaikinBook},
\begin{equation}
K = \left.-V \frac{\partial P}{\partial V}\right|_{V} = \left.\rho \frac{\partial P}{\partial \rho}\right|_{V},
\label{eq_PKV}
\end{equation}
characterizing the elastic response with respect to a volumetric (dilatational) strain fluctuation 
as shown in Fig.~\ref{fig_sketch}(d).
As discussed in the standard textbooks \cite{Callen,ChaikinBook}, the extensive variable is here
the volume $X=V = \langle \Vhat \rangle$ and the conjugated intensive variable the negative pressure 
$I=-P = - \langle \Phat \rangle$ where $\Vhat$ and $\Phat$ stand for the
instantaneous values of the volume and the pressure. With these notations Eq.~(\ref{eq_PKV}) is
obviously consistent with Eq.~(\ref{eq_FtoM}) or Eq.~(\ref{eq_HtoM}).
In analogy with $\Gmodgt$ for the shear modulus $G$, the compression modulus $K$ may be determined in 
an \NPgT-  or an \NPtT-ensemble using the linear regression relation
\begin{equation}
\KmodVP  \equiv \left.-V \la \dVhat \dPhat \ra / \la \dVhat^2 \ra\right|_{P}
\label{eq_KmodVP}
\end{equation}
where the index $vp$ indicates that both the measured instantaneous volume $\Vhat$ and pressure $\Phat$
have been used. The corresponding correlation coefficient is
\begin{equation}
\cmodVP \equiv \left. - \la \dVhat \dPhat \ra / \sqrt{\la \dVhat^2 \ra \la \dPhat^2 \ra}\right|_{P}.
\label{eq_cmodVP}
\end{equation}
Since from Eq.~(\ref{eq_dXdI}) we have
$\langle \dVhat \dPhat \rangle = - \kBT$ if $P$ is imposed,
this implies that $\KmodVP$ should be identical to the better known relation \cite{Callen}
\begin{equation}
\KmodVV \equiv \left.\kBT V/\la \dVhat^2 \ra\right|_{P}
\label{eq_KmodVV}
\end{equation}
which corresponds to the operational definition $\Gmodgg$ discussed above.
The compression modulus $K$ may of course also be determined using the pressure fluctuations
in systems of imposed volume $V$ as sketched in Fig.~\ref{fig_sketch}(c).
As (to our knowledge) first derived by Rowlinson \cite{RowlinsonBook},
the corresponding stress fluctuation formula reads
\begin{equation}
\KmodPP \equiv \etaA - \etaF \mbox{ with }
\etaF \equiv \beta V \la \dPhat^2 \ra
\label{eq_KmodPP}
\end{equation} 
being the fluctuation of the total normal pressure (corresponding to $\muF$)
and $\etaA$ the ``affine dilatational elasticity" (corresponding to $\muA$).
We shall give below a proper operational definition for $\etaA$.
Here we only note that due to the general Legendre transformation Eq.~(\ref{eq_dIdI}) for intensive variables
one knows immediately that
\begin{equation}
\left. \etaF\right|_{V} = \left. \etaF\right|_{P} - K.
\label{eq_etaFtransform}
\end{equation}
This implies by comparison with Eq.~(\ref{eq_KmodPP}) that
\begin{equation}
\etaA \stackrel{!}{=} \left. \etaF\right|_{P}
\label{eq_etaAetaFP}
\end{equation}
in analogy to Eq.~(\ref{eq_muAmuFtau}) for the shear,
i.e. $\etaA$ can be directly be determined from the pressure fluctuations
in \NPgT- and \NPtT-ensembles.
As before for the correlation coefficient $\cmodgt$, Eq.~(\ref{eq_cmodgt_muA}), 
it follows from Eq.~(\ref{eq_etaAetaFP}) that
\begin{equation}
\cmodVP = \sqrt{\KmodVV/\left.\etaF\right|_{P}} = \sqrt{K/\etaA}.
\label{eq_cmodVP_etaA}
\end{equation}
Since $\cmodVP \le 1$, the affine dilatational elasticity sets an upper bound to the compression modulus $K$.

\subsection{Rowlinson's stress fluctuation formula}
\label{app_Kstressfluctu}

\paragraph*{Introduction.}
We reformulate here briefly the derivation of the stress fluctuation relation $\KmodPP$ for the compression modulus $K$
of an isotropic system at imposed volume $V$ by Rowlinson \cite{RowlinsonBook}.
With $V(0)$ being the volume of the unperturbed simulation box 
we assume a small relative volume change 
\begin{equation}
\epsilon \equiv V(\epsilon)/V(0) -1 
\label{eq_epsilondef}
\end{equation}
with all coordinates equally deformed as, e.g.
\begin{equation}
x(0) \Rightarrow x(\epsilon) = x(0) (1+\epsilon)^{1/d}
\label{eq_strain_x_epsilon}
\end{equation}
for the $x$-coordinate in $d$-dimensions. (The argument $(0)$ denotes again the reference system.)
Using a similar rescaling trick as for the shear modulus $G$ in Sec.~\ref{theo_stressfluctu}, 
the interaction energy $\Uexs(\epsilon)$ of a configuration $s$ of the strained system is expressed in terms of 
the coordinates (state) of the unperturbed system and the explicit metric parameter $\epsilon$. 
Due to the volume change we have now also to take into account the ideal gas (kinetic) contributions.
 
\paragraph*{General excess contributions.}
We derive first the excess contribution $\Pex$ to the total pressure $P = \Pid + \Pex$
and the excess contribution $\Kex$ to the total compression modulus $K = \Kid + \Kex$ 
valid for an arbitrary conservative potential.
The general relations Eqs.~(\ref{eq_dlogZdgam}-\ref{eq_ddZddgam}) stated above for an imposed shear strain 
$\gamma$ still hold with the relative volume change $\epsilon$ replacing $\gamma$. Using Eq.~(\ref{eq_dZdgam})
it follows for the excess pressure that
\begin{eqnarray}
\Pex & = & -\frac{\partial \Fex(V)}{\partial V} = 
- \frac{1}{V(0)} \frac{\partial \Fex(\epsilon)}{\partial (1+\epsilon)}  \nonumber \\
     & = & \la \Pexhat \ra \mbox{ with } 
\Pexhat \equiv  -\frac{1}{V(0)} \left.\frac{\partial \Uexs^{\prime}(\epsilon)}{\partial \epsilon}\right|_{\epsilon=0}
\label{eq_Peshat_us}
\end{eqnarray}
{\em defining} the instantaneous excess pressure \cite{foot_Ihat}.
In the second step we have taken the limit $\epsilon \to 0$ and have dropped the argument $(0)$.
The average is again evaluated using the weights of the unperturbed system, Eq.~(\ref{eq_average}).
Using Eq.~(\ref{eq_ddlogZddgam}) and Eq.~(\ref{eq_ddZddgam}) one obtains for the compression modulus
\begin{eqnarray}
\Kex & = & V \frac{\partial^2 \Fex(V)}{\partial V^2} 
= \frac{1}{V(0)} \frac{\partial^2 \Fex(\epsilon) }{\partial (1+\epsilon)^2} \nonumber \\
      & = & \la \Uexs^{\prime\prime}(\epsilon)\ra/V - \beta V \la \delta \Pexhat^2 \ra
\label{eq_Kexstressfluctu}
\end{eqnarray}
which is again understood to be taken at $\epsilon =0$. 
The first term $\etaAex \equiv \la \Uexs^{\prime\prime}(\epsilon)\ra/V$ in Eq.~(\ref{eq_Kexstressfluctu}) corresponds 
to the change of the system energy assuming an affine strain transformation for all particle positions.
It gives the excess contribution $\etaAex$ to the total affine dilatational elasticity $\etaA = \etaAid + \etaAex$.
The second term corresponds to the excess contribution $\etaFex$ to the total normal pressure fluctuation 
$\etaF = \etaFid + \etaFex$ mentioned in Sec.~\ref{app_Kstrainfluctu}. 
It corrects the overprediction of the affine strain contribution $\etaAex$ \cite{Barrat06}.

\paragraph*{Ideal gas contribution.}
We remind that for an ideal gas the isothermal compression modulus $\Kid$ 
is given by the ideal gas pressure $\Pid = \kBT \rho$ and that the ideal gas pressure fluctuation 
$\etaFid \equiv \beta V \langle \delta \Pidhat^2 \rangle$ becomes $\left.\etaFid\right|_{V} = \Pid$
at constant volume \cite{ChaikinBook}. 
Interestingly, using again the general Legendre transform Eq.~(\ref{eq_dIdI}) one sees that
\begin{equation}
\left.\etaFid\right|_{P} = \left.\etaFid\right|_{V} + \Kid = 2 \Pid
\label{eq_etaFidP}
\end{equation}
for the ideal gas pressure fluctuations at imposed pressure.

Due the assumed additivity of the total system Hamiltonian, the stress fluctuation 
formula for the total compression modulus may now be written 
\begin{equation} 
\KmodPP  =  \Kid + \Kex = \Pid + \etaAex - \left.\etaFex\right|_V.
\label{eq_KconstantV}
\end{equation}
While this result is useful for constant-$V$ ensembles, especially if MC simulations are used,
it is yet not in a form exactly equivalent to the general Legendre transform Eq.~(\ref{eq_KmodPP}).
Since the ideal and excess pressure fluctuations are decoupled, $\langle \delta \Pidhat \delta \Pexhat \rangle =0$,
this implies $\etaF = \etaFid + \etaFex$. One may thus rewrite Eq.~(\ref{eq_KconstantV}) as $\KmodPP = \etaA - \etaF$ with
\begin{equation}
\etaA \equiv 2\Pid + \etaAex =  \left.\etaFid\right|_{P} + \etaAex 
\label{eq_etaAexid}
\end{equation}
where we have used Eq.~(\ref{eq_etaFidP}) in the second step.
Assuming Eq.~(\ref{eq_etaAexid}) assures that $\etaA$ becomes equivalent to $\left.\etaF\right|_{P}$.
This has the nice feature that not only the functional $\KmodPP$, Eq.~(\ref{eq_KmodPP}), should give 
the correct compression modulus $K$ in the \NVgT \ and the \NVtT \ ensembles, but also that $\KmodPP$
must vanish properly if the ``wrong" \NPgT \ and \NPgT \ ensembles are used, just as $\Gmodtt$ vanishes
for \NVtT- or \NPtT-ensembles. 

\paragraph*{Pair interactions.}
Up to now we have not taken advantage of the fact that the conservative interaction potential is assumed to be a pair potential, 
Eq.~(\ref{eq_Upairinteraction}). This will be used now to express $\Pex$ and $\etaAex$ in terms of simple 
(ensemble independent) two-point correlation functions.
Equation~(\ref{eq_strain_x_epsilon}) implies that the squared distance $r^2$ between two particles transforms as
\begin{equation}
r^2(0) \Rightarrow r^2(\epsilon) = r^2(0) (1+\epsilon)^{2/d}.
\label{eq_strain_r_epsilon}
\end{equation}
It follows that
\begin{eqnarray}
\frac{d r^2(\epsilon)}{d\epsilon} & \to & \frac{2}{d} r^2(0), \label{eq_strain_rone_epsilon} \\
\frac{d^2 r^2(\epsilon)}{d\epsilon^2} & \to & \frac{2(2-d)}{d^2} r^2(0) \label{eq_strain_rtwo_epsilon} 
\end{eqnarray}
where we have taken finally for both derivatives the limit $\epsilon \to 0$.
For the first two derivatives of a general function $f(r(\epsilon))$ with respect to $\epsilon$ this implies
\begin{eqnarray}
\frac{\partial f(r(\epsilon))}{\partial \epsilon} &  \to & \frac{1}{d} r f^{\prime}(r) \label{eq_strain_froneBeps} \\
\frac{\partial^2 f(r(\epsilon))}{\partial \epsilon^2} &  \to  &
\frac{1}{d^2} \left(r^2 f^{\prime\prime}(r) + r f^{\prime}(r) \right) \nonumber \\
 & - & \frac{1}{d} r f^{\prime}(r) \label{eq_strain_frtwoBeps}
\end{eqnarray}
for $\epsilon \to 0$ and dropping the argument $(0)$ on the right-hand sides.
According to the general result Eq.~(\ref{eq_Peshat_us}) for the normal pressure and using Eq.~(\ref{eq_strain_froneBeps})
one confirms for pair interactions the expected virial relation \cite{RowlinsonBook,AllenTildesleyBook}
\begin{equation}
\Pexhat = - \frac{1}{d V} \sum_l \rijl u^{\prime}(\rijl)
\label{eq_PexKirkwood}
\end{equation}
for the instantaneous excess pressure where the interactions between pairs of particles $i < j$ are again labeled
by an index $l$.
Using Eq.~(\ref{eq_strain_rtwo_epsilon}) the excess contribution to the affine dilatational elasticity becomes 
\begin{eqnarray}
\etaAex & = & \etaB + \Pex \mbox{ where } \label{eq_etaAex} \\
\etaB & \equiv & \frac{1}{d^2 V} \la \sum_l \rijl^2 u^{\prime\prime}(\rijl) + \underline{\rijl u^{\prime}(\rijl)} \ra 
\label{eq_etaB}
\end{eqnarray} 
is sometimes called the ``hypervirial" contribution to the compression modulus \cite{RowlinsonBook,AllenTildesleyBook}.
It corresponds to the first term indicated in Eq.~(\ref{eq_strain_frtwoBeps}). 
By comparing Eq.~(\ref{eq_etaB}) with Eq.~(\ref{eq_muB}) it is readily seen that for isotropic systems
\begin{equation}
\etaB =  \left(1+\frac{2}{d}\right) \muB - \frac{2}{d} \Pex,
\label{eq_etaBmuB}
\end{equation}
i.e. $\etaB$ and $\muB$ and, hence, the affine dilatational elasticity $\etaA$ and affine shear elasticity $\muA$ 
are closely related. (This justifies to call $\etaB$ a ``Born coefficient".)
Please note that Eq.~(\ref{eq_KmodPP}) or Eq.~(\ref{eq_KconstantV}) together with Eq.~(\ref{eq_etaAex}) 
are perfectly consistent with the relations stated in the literature 
\cite{RowlinsonBook,AllenTildesleyBook,FrenkelSmitBook,SBM11,SXM12,XWP12}.
We emphasize finally that, as discussed for the affine shear elasticity $\muA$ in Sec.~\ref{theo_stressfluctu},
it is inconsistent to neglect the explicit excess pressure contribution to $\etaA$ in Eq.~(\ref{eq_etaAex}) but to keep
the underlined term of the Born coefficient $\etaB$, Eq.~(\ref{eq_etaB}) \cite{foot_KLegendre}.

\subsection{Numerical findings}
\label{app_Knumerical}

\begin{figure}[t]
\centerline{\resizebox{1.0\columnwidth}{!}{\includegraphics*{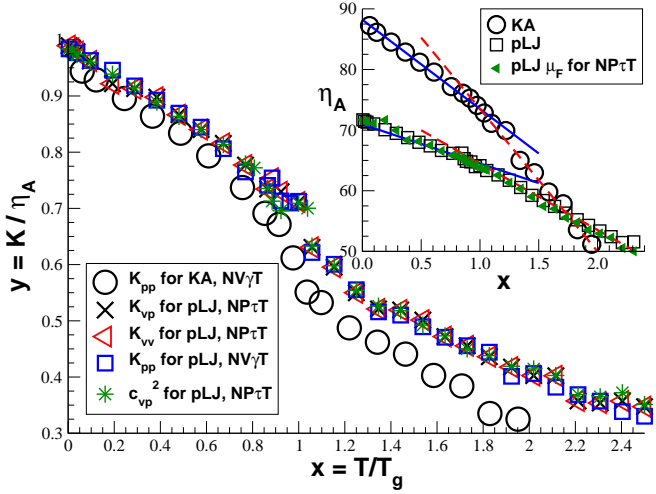}}}
\caption{Affine dilatational elasticity $\etaA(T)$ and compression modulus $K(T)$ for both models.
Inset: $\etaA$ {\em vs.} reduced temperature $x=T/\Tglass$.
The small triangles indicate $\etaF$ for the pLJ model using the \NPtT-ensemble confirming that Eq.~(\ref{eq_etaAetaFP}) holds.
Main panel:
The Rowlinson formula $\KmodPP$, Eq.~(\ref{eq_KconstantV}), for the \NVgT-ensemble is given for both the KA model
(spheres) and the pLJ model (squares). For the pLJ model we indicate in addition $\KmodVP$ (crosses), 
$\KmodVV$ (large triangles) and the rescaled correlation coefficient $\cmodVP^2$ (stars)
which collapses on the other pLJ data as suggested by Eq.~(\ref{eq_cmodVP_etaA}).
\label{fig_KT}
}
\end{figure}

\paragraph*{Affine dilatational elasticity.}
The affine dilatational elasticity $\etaA = \etaAid + \etaAex = 2 \Pid + \Pex + \etaB$ obtained for both models 
is presented in the inset of Fig.~\ref{fig_KT} as a function of the reduced temperature $x=T/\Tglass$. 
(The values may also be found in Table~\ref{tab_KA} and Table~\ref{tab_pLJ}.)
We remind that for determining the Born coefficient $\etaB$ impulsive corrections need to be 
considered as discussed at the end of Sec.~\ref{theo_trunc}.
As shown by the small filled triangles, $\etaA$ compares well with the total pressure fluctuation 
$\left.\etaF\right|_P = 2\Pid + \left.\etaFex\right|_P$ in an \NPtT-ensemble.
The observed data collapse is expected from Eq.~(\ref{eq_etaAetaFP}) and Eq.~(\ref{eq_etaFidP}).
As indicated by bold and dashed lines for, respectively, the low- and high-$T$ limit
$\etaA$ decreases essentially linearly with temperature. The coefficients of the lines (not shown) are consistent with
Eq.~(\ref{eq_etaBmuB}) relating $\etaA$ and $\muA$ and the fits given in Fig.~\ref{fig_muAmuF_T} for $\muA$.
We emphasize that the affine dilatational elasticity $\etaA$ (and thus $\muA$) is not only larger for the KA model 
but, more importantly that, is also increases more rapidly with decreasing $T$ than the pLJ model. 

\paragraph*{Compression modulus.}
The main panel of Fig.~\ref{fig_KT} shows the best long-time values of the compression modulus $K$ for both models.
The vertical axis is rescaled by $\etaA$. The rescaled stress fluctuation formula, Eq.~(\ref{eq_KconstantV}), 
for the \NVgT-ensemble is given for both models. For the pLJ model we also indicate $\KmodVP$ (crosses) and 
$\KmodVV$ (large triangles) for the \NPtT-ensemble. No rescaling with $\etaA$ is necessary for the squared 
correlation coefficient $\cmodVP^2$ (stars) also presented. A perfect collapse of all pLJ data is found confirming the
equivalence of all operational definitions used. 
For both models we observe a strong increase of $K/\etaA$ around $x \approx 1$ with decreasing temperature.
Interestingly, the rescaled data of both models is similar but not identical. The mentioned increase
around $x \approx 1$ is clearly more pronounced for the KA model. Note that if the unscaled $K$ is directly 
plotted against $T$, this effect becomes even stronger (not shown) consistently with the scaling of $\etaA$ presented in the inset.
We note finally that for very low temperatures we observe for both models that the ratio $K/\etaA = \cmodgt^2$
approaches unity, i.e. the compression modulus is essentially dominated by the affine response.
We remind that this is different for the reduced shear modulus for both models which is observed
to approach $G/\muA \approx 1/2$ in the low-$T$ limit (Fig.~\ref{fig_GT_rescal}), 
i.e. the shear stress fluctuations $\muFgam$ are more relevant than the pressure fluctuations $\left.\etaF\right|_{V}$ in this limit.

\begin{figure}[b]
\centerline{\resizebox{1.0\columnwidth}{!}{\includegraphics*{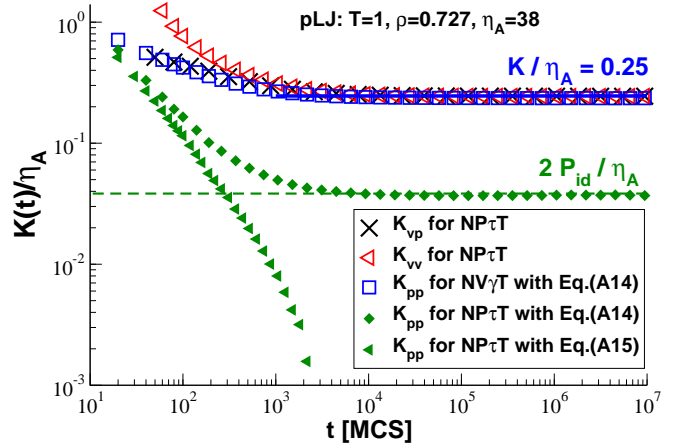}}}
\caption{Rescaled compression modulus $K(t)/\etaA$ {\em vs.} sampling time $t$ in MCS in the 
high-$T$ liquid regime for pLJ particles at $P=2$ and $T=1$ comparing $\KmodVP$, $\KmodVV$
and $\KmodPP$ in different ensembles as indicated. The filled symbols refer to $\KmodPP$
obtained in the ``wrong" \NPtT-ensemble. 
\label{fig_Kt_Thigh}
}
\end{figure}

\paragraph*{Sampling time dependence.}
For the pLJ model we present in Fig.~\ref{fig_Kt_Thigh} the sampling time dependence
of various operation definitions of $K$ for the liquid limit at temperature $T=1.0$.
The horizontal axis indicates the sampling time $t$ in units of MCS, the vertical axis 
is made dimensionless using $\etaA$ as natural scale.
As shown for $\KmodVP$ and $\KmodVV$ obtained in the \NPtT-ensemble and for $\KmodPP$ 
obtained using Eq.~(\ref{eq_KconstantV}) in the \NVgT-ensemble, all operational definitions
applied in their natural ensemble converge {\em rapidly} within $t \approx 10^3$ MCS to the same plateau. 
(This is a general finding for all temperatures.)
The plateau value for $T=1.0$ is indicated by the bold solid line.
Note that $K/\etaA$ is smaller unity as expected from the discussion at the end of Sec.~\ref{app_Kstrainfluctu}.

The small filled symbols refer to $\KmodPP$ computed in the ``wrong" \NPtT-ensemble.
The small diamonds have been computed using directly Eq.~(\ref{eq_KconstantV}).  
As indicated by the dashed line, this yields $\KmodPP \approx 2\Pid$ for large times.
Only if we compute $\KmodPP = \etaA - \etaF$ using Eq.~(\ref{eq_etaAexid}), 
it does vanish properly as shown by the small triangles.  
This shows numerically that Eq.~(\ref{eq_KconstantV}), albeit perfectly fine for constant-$V$ ensembles, 
is just not the correct formulation allowing to make manifest the general Legendre transformation 
Eq.~(\ref{eq_etaFtransform}). This is, however, achieved using Eq.~(\ref{eq_etaAexid}).
This finding holds rigorously for all temperatures as can be seen from the scaling of the 
small triangles in the inset of Fig.~\ref{fig_KT}.
%


\begin{thebibliography}{60}
\bibitem[{foo({\natexlab{a}})}]{foot_dgamdtau}
\bibinfo{note}{We have computed the static correlation function $\beta
  V\left.\la \dgamhat \dtauhat \ra\right|_{\tau}$ for a broad range of
  temperatures and pressures for both \NVtT- and \NPtT-ensembles. Converging
  rapidly to unity, this confirms that Eq.~(\ref{eq_dgamdtau}) holds to a high
  accuracy well up into the liquid regime.}
\bibitem[{foo({\natexlab{c}})}]{foot_Ihat}
\bibinfo{note}{There is considerable freedom for defining $\Ihat$ as long as
  the average $I$ does not change. The experimentally natural choice for
  $\Ihat=\tauhat$ is given by the instantaneous horizontal force (per surface)
  on the top wall in Fig.~\ref{fig_sketch}(a). For theory and simulation,
  Eq.~(\ref{eq_tauhat_us}) or Kirkwood's formula,
  Eq.~(\ref{eq_tauhatKirkwood}), for pair interactions are more suitable for
  $\tauhat$, but by no means unique.}
\bibitem[{foo({\natexlab{d}})}]{foot_Mbound}
\bibinfo{note}{Stability requires, of course, the convexity of the free energy,
  i.e. $M > 0$. We assume here that the formalism can be applied for
  arbitrarily small $M$.}
\bibitem[{foo({\natexlab{f}})}]{foot_metricchange}
\bibinfo{note}{Imposing an affine strain corresponds merely to changing the
  interaction Hamiltonian for the same state of the system. In practice this is
  done by changing the well-known $h$-matrix \cite{FrenkelSmitBook} in our
  computer simulations.}
\bibitem[{foo({\natexlab{g}})}]{foot_localonly}
\bibinfo{note}{The inefficiency of local MC moves at low temperatures can
  easily be understood by considering a bead-spring model
  \cite{LandauBinderBook} of a very stiff polymer chain for which the
  relaxation time $\taurelax(M)$ must increase exponentially with the molecular
  mass $M$ if only local MC jumps are used. A more efficient power law scaling
  of $\taurelax(M)$ is obtained if global collective MC moves associated with
  the translational motion of the chain center of mass and the orientation of
  the chain end-to-end vector are added.}
\bibitem[{foo({\natexlab{h}})}]{foot_adaptive}
\bibinfo{note}{In principle, it would be better to change the maximum amplitude
  of the plane waves adaptively in order to fix an acceptance rate
  $A(q,T)=const$ for all wavevectors and temperatures. This has yet not been
  done for the present study. As shown in Sec.~\ref{res_dynmat} for the
  sampling of fixed networks constructed from the dynamical matrix
  corresponding to the colloidal glass, some systems have also been computed
  successfully only using global moves. It is thus not stictly necessary to
  include the computationally expensive local MC moves. In any case, the global
  plane-wave MC algorithm requires a more systematic theoretical and numerical
  analysis to improve its computational efficiency.}
\bibitem[{foo({\natexlab{i}})}]{foot_maxvalues}
\bibinfo{note}{For the temperature quench of the configurations sufficiently
  small values for $\delta \gamma_\mathrm{max}$ for a shear strain altering MC
  move and for $\delta V_\mathrm{max}$ for a volume change have been chosen
  such that the acceptance rate of both MC steps remains reasonable at the
  lowest temperatures ($T < 0.1$). As for the global plane-wave MC moves, it
  might better to change these maximum amplitudes adaptively while the quench
  is performed to keep fixed constant acceptance rates.}
\bibitem[{foo({\natexlab{j}})}]{foot_dynamics}
\bibinfo{note}{Similar values for $\Tglass$ have also been obtained for both
  models following the standard diagnostics described in the literature
  \cite{Barrat88,Kob95} of various dynamic properties such as the scaling of
  the mean-square displacements, the self-diffusion coefficient or the Fourier
  transformed density autocorrelation function in the low-wavevector limit.
  From the analysis of these dynamical properties it can be seen that compared
  to the KA model the pLJ model is a much more gradual glass-former
  qualitatively similar to the purely repulsive soft-sphere model studied in
  Ref.~\cite{Barrat88}.}
\bibitem[{foo({\natexlab{k}})}]{foot_KLegendre}
\bibinfo{note}{As for the shear stress fluctuations at constant $\tau$ in
  Sec.~\ref{theo_Legendre}, it is possible to demonstrate by integration by
  parts, Eq.~(\ref{eq_intbyparts}), that Eq.~(\ref{eq_etaAetaFP}) holds with
  $\etaA$ being given by Eq.~(\ref{eq_etaAexid}) and Eq.~(\ref{eq_etaB}). Using
  the Legendre transform Eq.~(\ref{eq_etaFtransform}) one thus immediately
  obtains Rowlinson's pressure fluctuation formula, Eq.~(\ref{eq_KmodPP}).}
\end{thebibliography}

\end{document}